\documentclass[11pt]{article}
\newcommand{\lambdabar}{{\mkern0.75mu\mathchar '26\mkern -9.75mu\lambda}}
\begin{document}
\title
{Response of simple quantum systems to different polarizations of gravitational waves in noncommutative phase-space }
\author{
{\bf {\normalsize Sunandan Gangopadhyay}$^{a,c}
$\thanks{sunandan.gangopadhyay@gmail.com}},
{\bf {\normalsize Anirban Saha}
$^{a,c}$\thanks{anirban@iucaa.ernet.in}},
{\bf {\normalsize Swarup Saha}$^{b,a}$\thanks{saha18swarup@gmail.com}}\\[0.2cm]
$^{a}$ {\normalsize Department of Physics, West Bengal State University,
Barasat, Kolkata 700126, India}\\[0.2cm]
$^{b}$ {\normalsize Department of Radiotherapy and Nuclear Medicine, Barasat Cancer Research and Welfare Center,}\\
{\normalsize Barasat, India}\\[0.2cm]
$^{c}${\normalsize Visiting Associate in Inter University Centre for Astronomy $\&$ Astrophysics,}\\
{\normalsize Pune, India}\\[0.3cm]
}
\date{}

\maketitle

\begin{abstract}
\noindent Owing to the extreme smallness of any noncommutative scale that may exist in nature, both in the spatial and momentum sector of the quantum phase-space, a credible possibility of their detection lies in the present day gravitational wave detector set-ups, which effectively detects the relative length-scale variations ${\cal{O}}\left[10^{-23} \right]$. With this motivation, we have considered how a free particle and harmonic oscillator in a quantum domain will respond to linearly and circularly polarized gravitational waves if the given phase-space has a noncommutative structure. The results show resonance behaviour in the responses of both free particle and HO systems to GW with both kind of polarizations. We critically analyze all the responses, and their implications in possible detection of noncommutativity. We use the currently available upper-bound estimates on various noncommutative parameters to anticipate the relative size of various response terms. We also argue how the quantum harmonic oscillator system we considered here can be very relevant in context of the resonant bar detectors of GW which are already operational currently.
   
\end{abstract}

\maketitle
\section{Introduction}
\noindent At the Planck scale the space-time is thought to have a granular structure much like the phase-space of quantum mechanics (QM). This is owing to the uncertainty induced in spatial coordinates due to sharp localization of events in space \cite{Dop1, Dop2, Alu}. This granularity can be realized theoretically by describing the space-time with a set of coordinates $x^{\mu}, \left( \mu = 0,1,2,3 \right)$ following a noncommutative (NC) algebra $\left[x^{\mu}, x^{\nu}\right] = i \theta^{\mu \nu}$, where $\theta^{\mu \nu}$ is a constant anti-symmetric tensor of second rank. This space-time is referred to as canonical NC space-time \cite{snyder}. 
Theories defined on such spaces are generically called NC theories \cite{sw, szabo, ref4}. One can further extend the NC space to a more general NC phase-space\footnote{Since we only consider the NC quantum mechanical phase-space we choose to ignore noncommutativity among the spatial coordinates and time.} \cite{momentum_NC1, momentum_NC2, momentum_NC3, bert0, samanta} with both the canonical pairs following the algebra: 
\begin{eqnarray}
\left[{\hat x}_{i}, {\hat p}_{j}\right] = i\tilde{h}\delta_{ij} \>, \quad 
\left[{\hat x}_{i}, {\hat x}_{j}\right] = i \theta \epsilon_{ij} \>,\quad 
\left[{\hat p}_{i}, {\hat p}_{j}\right] =i \bar{\theta} \epsilon_{ij} \>
\label{e9a}
\end{eqnarray}
where $i,j = 1, 2; \tilde{h}= \hbar\left({1} + \frac{\theta\bar{\theta}}{4\hbar^2}\right)$ is the modified Planck's constant; $\theta$ and $\bar{\theta}$ are spatial and momentum NC parameters respectively and $\epsilon_{ij}= -\epsilon_{ji}, (\epsilon_{12}=1)$ is the anisymmetric tensor in two dimension\footnote{For our purpose we can confine our attention to the four-dimensional NC phase-space}. The main argument for considering a NC phase-space is that noncommutativity between momenta arises naturally as a consequence of noncommutativity between coordinates, as momenta are defined as the partial derivatives of the action with respect to the coordinates \cite{momentum_NC4}. Theories defined over a noncommutative phase-space have also been furnished in \cite{giri, bensg, dulat} in context of the NC harmonic oscillator and NC Lorentz transformations. 

The mathematical complexities aside, the biggest challenge in such NC theories is to identify experimentally detectable effects of noncommutativity owing to the extreme smallness of the NC parameters $\theta $ and $\bar{\theta} $ appearing in the algebra (\ref{e9a}). Though such effects may only appear near the string/Planckian scale, it is hoped that some low energy relics may exist and their phenomenological consequences are currently being explored at the level of quantum mechanics \cite{Nair, bcsgas, pmh, bert0, ani, galileo, mezin, hazra, gov}. Typical low energy non-accelerator experiments are the Lamb-shift \cite{cst}, and clock-comparison experiments \cite{carol} where the upperbound on the value of the canonical NC parameter was found to be $\theta\leq \left(10 {\rm TeV}\right)^{-2}$ which corresponds to $4 \times 10^{-40} {\rm m}^{2}$ for $\hbar$$=$$c$$=$$1$. On the other hand such upperbounds on the momentum NC parameter \cite{bert0, samanta} is $\bar{\theta} \leq 2.32 \times 10^{-61} {\rm kg^{2}m^{2}sec^{-2}}$ and time-space NC parameter \cite{ani, galileo} is $\theta^{0i} \leq 9.51\times 10^{-18} {\rm m}^{2}$ which are shown to be mutually consistent in \cite{ani}. 
These upperbounds correspond to the length scale range $\sim 10^{-20} {\rm m} $. 
Lookings at these bounds one can not help but notice that at present the only potential possibility of finding the NC signature experimentally is in the high-precision gravitational wave detection experiments like LIGO \cite{abramovici}, VIRGO \cite{caron}, GEO \cite{luck} and TAMA \cite{ando}. Indeed, the current Advanced LIGO \cite{Adv_Ligo} detectors has reached a sensitivity where one can detect a length-variation of the order of $\frac{\delta L}{L} \sim 10^{-23}/\sqrt{{\rm{Hz}}} $ or better. Various resonant bar detectors are also not much behind \cite{bar_2, bar_detectors_1, bar_detectors_2, bar_detectors_3, bar_detectors_4, bar_detectors_5}.
This roughly means that these modern GW interferometers are capable of monitoring the (relative) positions of their test masses with an accuracy of order $10^{-20}$m. Also, GW detectors based on matter-wave interferometry \cite{atom_int, atom_int1, atom_int2} have been suggested recently which claim similar strain sensitivity. These interferometers are thus ideally suited to monitor the fuzzyness introduced in distance measurements between test masses. If the space we live in indeed has a NC phase-space structure, the concept of distance is then fundamentally fuzzy. Thus even in the idealized situation where all classical and ordinary quantum noise sources are completely eliminated, the read-out of a GW detector would still pick up the noise of phase-space noncommutativity \cite{nature, cam}. 

Since NC phase-space structure is inherently quantum mechanical in nature, a quantum mechanical theory of the GW-matter interaction in an NC phase-space would be necessary to predict the possible NC effects, noise sources or otherwise, in the GW detector read-outs. In this endeavour we have recently adopted a systematic approach \cite{speli} to study the effect of linearly polarized GW(s) in the long-wavelength and low velocity limit on the test matter, e.g., free particle and a harmonic oscillator (HO) \cite{ncgw1, ncgw2} in a space with NC coordinates. These studies show that while the spatial noncommutativity may not affect the response of a free particle to GW in a detectable manner, the response of a HO system to GW is significantly affected. Specifically, the spatial noncommutativity introduces a characteristic frequency into the HO system that shifts the resonance point in presence of GW, moreover it also creates a new resonance between the natural frequency of the HO and itself \cite{ncgw2}. In \cite{ncgw3} we have shown that when the noncommutative structure is generalized from the septial sector to the whole phase-space, even the response of the free particle to linearly polarized GW is significantly altered. The noncommutativity in the momentum sector brings in an oscillatory nature in the dynamical evolution of the free particle with a frequency characterized by the momentum NC parameter. Also in \cite{ncgw4} we have shown that the free particle and HO, in presence of only spatial noncommutativity, respond to a circularly polarized GW with the same basic features as they do to linearly polarized ones. Accounting for the circular polarization is non-trivial primarily because, unlike the linearly polarized case, here the polarization vectors evolve in time which makes the algorithm for computing the time evolution of the responding matter different from the former case.

In view of these results one can immediately identify that three very pertinent questions emerge:
\begin{enumerate}
	\item
Do the matter system retain the basic similarities in their responses to both linear and circularly polarized GW (as has been found in the results of \cite{ncgw4} mentioned above) when one generalizes from spatial noncommutativity to phase-space noncommutativity?
	\item
Since the momentum sector of noncommutativity is known to introduce a novel frequency characterizing a fundamental momentum scale into the free particle system \cite{ncgw3}, making its time evolution periodic, how it will affect the HO system which already has a natural oscillation that can resonate with periodic GW signals? 
	\item
Are the effects of the spatial and momentum sector of the NC phase space distinguishable? Posed in a different form, if a NC signature is indeed detected in the response of free particle/HO to linearly/circularly polarized GW, can one tell from which sector of the NC phase-space algebra did it originate?
\end{enumerate}
We answer all these questions in the present paper by constructing a general quantum mechanical description in the NC phase-space where both coordinate and momentum are assumed to follow noncommutative algebra (\ref{e9a}), with both linear and circular polarization of GW interacting with free particle/ HO systems. We shall carry out the analysis for the linearly and circularly polarized GW signals separately since the algorithm to compute time evolution for the two cases are different.

  This paper is organized as follows. In the next section we describe the general methodology of our work. This involves obtaining a quantum mechanical description of a simple matter system interacting with GW in the NC phase-space and computing its subsequent time-evolution. This will lay the theoretical foundation of the paper and also fix the notations. In section 3, we compute the dynamics of a quantum particle which is otherwise free except for its coupling to circularly polarized GW, in NC phase space. A comparison of this result with our earlier work \cite{ncgw3} where exclusively linearly polarized GWs was considered will follow. In section 4 we elaborate on the quantum dynamics of a harmonic oscillator coupled to GW, again in the NC phase-space, and discuss the results. Here interaction with both linearly as well as circularly polarized GW will be worked out, the results will be conclude with a discussion in section 5.  
\section{Methodology}
We start by noting that in the proper detector frame the geodesic deviation equation for a particle of mass $m$ subject to linearized GW takes \cite{Magg} the form of a Newton's force equation. Specifically, in a potential $V\left(x\right)$ the equation is
\begin{equation}
m \ddot{{x}} ^{j}= - m{R^j}_{0,k0} {x}^{k} - \partial_{j} V
\label{e5}
\end{equation}
where dot denotes derivative with respect to the coordinate time of the proper detector frame\footnote{It is the same as it's proper time to first order in the metric perturbation.}, ${x}^{j}$ is the proper distance of 
the particle from the origin and ${R^j}_{0,k0}$ are the relevant components of the curvature tensor
in terms of the metric perturbation  $h_{\mu\nu}$ defined by\footnote{As is usual, latin indices run from $1-3$. Also $;$ denotes covariant derivatives.}   
\begin{eqnarray}
g_{\mu\nu} = \eta_{\mu\nu} + h_{\mu\nu}; \, |h_{\mu\nu}|<<1
\label{metric_perturbation}
\end{eqnarray}
on the flat Minkowski background $\eta_{\mu\nu}$.

The 
gauge-choice
$\left( h_{0\mu} = 0,  h_{\mu\nu;}{}^{\mu} =0, h_\mu^\mu=0 \right)$
that removes the unphysical degrees of freedom (DOF) and renders the GW transverse and traceless, has been made, so that only non-trivial components of the curvature tensor 
${R^j}_{0,k0} = -\ddot{h}_{jk}/2$
appear in eq.(\ref{e5}). The two physical DOF brought out by this choice are referred as the $\times$ and $+$ polarizations of GW and considering $z$ to be the propagation direction, the surviving components of the $2\times 2$ matrix $h_{jk}$ in the transverse 
plane, $h_{11} = -h_{22}$ and $ h_{12} = h_{21}$ represent these states respectively.

Note that, eq.(\ref{e5}) can be used as long as the spatial velocities involved are non-relativistic and $|{x}^{j}|$ is much smaller than 
the reduced wavelength $\frac{\lambda}{2\pi}$ of GW. These conditions are collectively referred as the \textit{small-velocity and long wavelength limit} and met by resonant bar-detectors and earth bound interferometric detectors\footnote{Note that these conditions are not satisfied by the proposed space-borne interferometer LISA  or by the Doppler tracking of spacecraft.} with the origin of the coordinate system centered at the detector. This also ensures that in a plane-wave expansion of GW, 
$h_{jk} = \int  (A_{jk} e^{ikx} + A^{*}_{jk} e^{- ikx})
d^{3}k/\left(2 \pi\right)^{3},$
the spatial part $e^{i \vec{k}.\vec{x}} \approx 1$ all over the detector site. Thus our only concern is the time-dependent part of the GW. If the polarization information contained in $A_{jk}$ is expressed in terms of the Pauli spin matrices, $h_{jk}$ takes the most convenient form 
\begin{equation}
h_{jk} \left(t\right) = 2f \left(\varepsilon_{\times}\sigma^1_{jk} + \varepsilon_{+}\sigma^3_{jk}\right) 
\label{e13}
\end{equation}
where $2f$ is the amplitude of the GW and $\left( \varepsilon_{\times}, \varepsilon_{+} \right)$ are the two possible polarization states of the GW satisfying the condition $\varepsilon_{\times}^2+\varepsilon_{+}^2 = 1$ for all $t$. For linearly polarized  GW the frequency $\Omega$ is contained in the time-dependent amplitude $2f(t)$ whereas for circularly polarized GW the time-dependent polarization states $\left( \varepsilon_{\times} \left(t \right), \varepsilon_{+} \left( t \right) \right)$ contains the frequency $\Omega$.
\subsection{Constructing the quantum system interacting with Gravitational waves in NC phase-space}
In order to obtain a quantum mechanical description of the GW-matter interaction we first obtain the appropriate classical Hamiltonian. This can be simply done by writing down the Lagrangian for the system (\ref{e5}), upto a total time- derivative term as 
\begin{equation}
{\cal L} = \frac{1}{2} m\dot {x}^2 - m{\Gamma^j}_{0k}\dot {x}_{j} {x}^{k}  -  V\left(x\right)
\label{e8}
\end{equation}
where ${R^j}_{0,k0} = - \frac{d \Gamma^j_{0k}}{d t}  = -\ddot{h}_{jk}/2  $.
Computing the canonical momentum ${p}_{j} = m\dot {x}_{j} - m \Gamma^j_{0k} {x}^{k}$  corresponding to ${x}_{j}$ we write the Hamitonian 
\begin{equation}
{H} = \frac{1}{2m}\left({p}_{j} + m \Gamma^j_{0k} {x}^{k}\right)^2 + V\left(x\right)   \>.
\label{e9}
\end{equation}
Once we have the classical Hamiltonian we can have the NCQM description of the system simply by elevating the phase-space variables $\left( x^{j}, p_{j} \right)$ to operators $\left( {\hat x}^{j}, {\hat p}_{j} \right)$ and imposing the NC Heisenberg algebra (\ref{e9a}). 
Note that this algebra can be mapped \cite{cst, stern} to the standard $\left( \theta, \bar{\theta}  \right) = \left(0, 0 \right)$ Heisenberg algebra spanned by the operators $\hat{X}_{i}$ and $\hat{P}_{j}$ of the ordinary QM  through the transformation equations
\begin{eqnarray}
{\hat x}_{i} = {\hat X}_{i} - \frac{1}{2 \hbar} \theta \epsilon_{ij} {\hat P}_{j}\>, \quad {\hat p}_{i} = {\hat P}_{i} + \frac{1}{2 \hbar} 
\bar{\theta} \epsilon_{ij} {\hat X}_{j}\>\>.
\label{e9b}
\end{eqnarray}
Writing the NCQM version of eq.(\ref{e9}) and employing the map (\ref{e9b}) we obtain the effective Hamiltonian 
\begin{eqnarray}
\hat{H} = \frac{ {\hat P}_{j}{}^{2}}{2m} + \Gamma^j_{0k} {\hat X}_{j} {\hat P}_{k} + \frac{\bar{\theta}}{2 m\hbar} \epsilon_{jm} {\hat X}^{j} {\hat P}_{m}  
- \frac{\theta }{2 \hbar} \epsilon_{jm}  {\hat P}_{m} {\hat P}_{k}  \Gamma^j_{0k} \>
+\frac{{\bar{\theta}} }{2 \hbar} \epsilon_{jm} {\hat X}_{m} {\hat X}_{k}  \Gamma^j_{0k} + V\left({\hat X}_{i} - \frac{1}{2 \hbar} \theta \epsilon_{ij} {\hat P}_{j}\right)
\label{e12}
\end{eqnarray}
which gives the equivalent description of the noncommutative system (\ref{e9}) in terms of the commutative operators ${\hat X}_{i}$ and ${\hat P}_{j}$. Since these operators admit the standard Heisenberg algebra, the rules of ordinary QM applies to (\ref{e12}). Also note that it has been demonstrated in various formulations of NC general relativity \cite{grav, banerjee11} that any NC correction in the gravity sector is second order in the NC parameter. Therefore, in a first order theory in NC space, the GW remains unaltered by NC effects.
\subsection{Tracking the time-evolution}
To study the time evolution of the system we introduce the raising and lowering operators 
\begin{eqnarray}
\hat{X}_j = \left({\hbar\over 2m\varpi}\right)^{1/2}
\left(a_j+a_j^\dagger\right)\quad;\quad
\hat{P}_j = -i\left({\hbar m\varpi\over 2}\right)^{1/2} 
\left(a_j-a_j^\dagger\right)
\label{xp}
\end{eqnarray}
For the free particle case, $V\left( x \right) = 0$, the frequency $\varpi$ is determined from the initial uncertainty in either the position or the momentum of the particle \cite{speli} whereas for the harmonic oscillator case $V\left( x \right) = \frac{1}{2} m \varpi^{2} x_{j}{}^2$, $\varpi$ is identified with the natural frequency of the oscillator. Expressing the commutative equivalent effective Hamiltonian (\ref{e12}) in terms of $(a_j, a_j^\dagger)$ one can easily compute the time evolution of the system by
\begin{eqnarray}
\frac{da_{j}(t)}{dt} = \frac{1}{i\hbar}\left[\hat{H} , a_{j} \right]
\label{evolution}
\end{eqnarray}
using the the algebra satisfied by the raising and lowering operators 
\begin{eqnarray}
\left[a_j(t), a^\dagger_k(t)\right] = \delta_{jk}\quad;\quad
\left[a_j(t), a_k(t)\right] = 0 = \left[a^\dagger_j(t), a^\dagger_k(t)\right].
\label{e18}
\end{eqnarray}
We further employ the time dependent Bogoliubov transformation which relate the operators $a_j(t)$ and $a_j^{\dagger}(t)$ with their initial value at time $t=0$ 
\begin{eqnarray}
a_j(t) = u_{jk}(t) a_k(0) + v_{jk}(t)a^\dagger_k(0)\quad;\quad 
a_j^\dagger(t) = a_k^\dagger(0)\bar u_{kj}(t)  + a_k(0)\bar v_{kj}(t)\>
\label{e19}
\end{eqnarray}
so that the time evolution of the system can be cast in terms of the generalized Bogoliubov coefficients $u_{jk}$ and $v_{jk}$ which are $2\times 2$ complex matrices in the $x-y$ plane. 
Due to eq.~$(\ref{e18})$, they must satisfy
\begin{equation}
uv^{T}=u^{T}v\>,\> u u^\dagger - v v^\dagger = I,
\label{e20}
\end{equation}
written in matrix form where $T$ denotes transpose, $\dagger$ denotes complex conjugate transpose and $I$ is the identity matrix. 
Since $a_j(t = 0) = a_j(0)$, equation (\ref{e19}) imposes the boundary conditions  
\begin{eqnarray}
u_{jk}(0) =  I  \quad;\quad  v_{jk}(0) = 0~.
\label{bc0}
\end{eqnarray}
on $u_{jk}(t)$ and $v_{jk}(t)$.
The final form of the equations of motion are writen in terms of a pair of  $2\times 2$ matrices 
\begin{eqnarray}
\zeta_{jk} = u_{jk} - v_{jk}{}^\dagger; \, \xi_{jk} = u_{jk} + v_{jk}{}^\dagger
\label{matrix_pair}
\end{eqnarray}
which are in the $x-y$ plane. 
\subsubsection{Case of linearly polarized GW}
The linearly polarized GW can be expressed in terms of the Pauli spin matrices (see equation (\ref{e13})) as
\begin{equation}
h_{jk} \left(t\right) = 2f(t) \left(\varepsilon_{\times}\sigma^1_{jk} 
+ \varepsilon_{+}\sigma^3_{jk}\right) 
\label{e13a}
\end{equation}
To set a suitable boundary condition we shall assume that the GW hits the system at $t=0$ so that 
\begin{equation}
f(t)=0 \>, \quad {\rm for} \ t \le 0.
\label{bc}
\end{equation}
Since any $2\times 2$ complex matrix can be written as a linear combination of the Pauli spin matrices and identity matrix, we also express the matrices $\zeta_{jk}$ and $\xi_{jk}$ as
\begin{eqnarray}
\zeta_{jk}\left(t \right) &=& A I_{jk} + B_{1}\sigma^{1}_{jk}  
+ B_{2}\sigma^{2}_{jk} + B_{3}\sigma^{3}_{jk} 
\label{form2}\\
\xi_{jk} \left(t \right) &=&  C I_{jk} + D_{1} \sigma^{1}_{jk} 
+ D_{2} \sigma^{2}_{jk} + D_{3} \sigma^{3}_{jk}~.
\label{form1}  
\end{eqnarray} 
so that the equations of motion will reduce to a set of coupled first order differential equations in $A, B_{1}, B_{2}, B_{3}, C, D_{1}, D_{2}, D_{3}$. Noting that $|f(t)|<<1$, these equations can be solved
iteratively about their $f(t)=0$ solution.
\subsubsection{Case of circularly polarized GW}
In case of circularly polarized GW the situation is slightly different. Here the GW $h_{jk}$ can be written as 
\begin{equation}
h_{jk} \left(t\right) = 2f_{0}\left(\varepsilon_{1}(t)\sigma^1_{jk} + \varepsilon_{3}(t)\sigma^3_{jk}\right) 
\label{e13}
\end{equation}
where $2f_{0}$ is the constant amplitude of the GW and the polarization vectors $\varepsilon_{1}(t)$ and $\varepsilon_{3}(t)$
evolve according to 
\begin{equation}
\frac{d\epsilon_{3}(t)}{dt} = \Omega
\epsilon_{1}(t)~,~\frac{d\epsilon_{1}(t)}{dt} = - \Omega\epsilon_{3}(t)
\label{e23}
\end{equation}
with $\Omega$ being a constant frequency. In this case we proceed to solve the equations of motion as follows. First we remember that any $2\times 2$ complex matrix $M$ can be written as a linear combination of the Pauli spin matrices and identity matrix as
\begin{equation}
M = \theta_0 I + \theta_A\sigma^A
\label{e23s}
\end{equation} 
where $\theta_0$ and $\theta_A$ are complex numbers. Next, considering $\left(\theta_{A}\right),A=1,2,3$  as being a vector in a three dimensional complex space.The polarization states of the GW can also be represented as a vector $\vec\varepsilon$ in this space. Considering $\vec\varepsilon$, $\dot{\vec\varepsilon}$ and $\vec\varepsilon\times\dot{\vec\varepsilon}$ are mutually orthogonal and thus form a natural directional triad, we can choose the trio as the coordinate axis for this space. Hence, we expess the pair of matrices $\left( \chi, \xi \right)$ as
\begin{equation}
\chi = A I + B\vec\varepsilon\cdot\vec\sigma +
       C\frac{\dot{\vec\varepsilon}\cdot\vec\sigma}{\Omega} +
       Di\frac{\vec\varepsilon\times
       \dot{\vec\varepsilon}}{\Omega}\cdot\vec\sigma\>,
\label{e33}
\end{equation}
\begin{equation}
\xi =  E I + F\vec\varepsilon\cdot\vec\sigma +
       G\frac{\dot{\vec\varepsilon}\cdot\vec\sigma}{\Omega} +
       Hi\frac{\vec\varepsilon\times
       \dot{\vec\varepsilon}}{\Omega}\cdot\vec\sigma\>,
\label{e43}
\end{equation}
where $A$, $B$, $C$, $D$,$E$,$F$,$G$,$H$ can be complex functions. We thus reduce our equations of motion to a set of first order differential equations for these complex fuctions which will be solved iteratively about the $f_{0}=0$ solution upto first order in the GW amplitude. In the next two sections we shall consider the response of a free particle and harmonic oscillator to GW respectively in a NC phase-space. 
\section{Free particle in NC phase space interacting with circularly polarized gravitational wave}
 We consider a quantum mechanical free particle interacting with circularly polarized GWs in NC phase-space that obeys algebra (\ref{e9a}). In this case  the potential is $V = 0$ and  in terms of the raising and lowering operators (\ref{xp}) the commutative equivalent of the NC phase-space Hamiltonian (\ref{e12}) takes the form
 \begin{eqnarray}
\hat{H} &=& \frac {\hbar\varpi}{4}\left(2 a_j^\dagger a_j + 1 -a_j^2 - a_j^{\dagger 2}\right) - \frac{i\hbar}{4} \dot h_{jk}\left(a_j a_k - a_j^\dagger a_k^\dagger\right)+ \frac {i\bar{\theta}}{4m}\epsilon_{jk}a_j^\dagger a_k  \nonumber\\
&&+\frac{m\varpi\theta}{8}\epsilon_{jm}\dot{h}_{jk}(a_m a_k -a_m a_{k}^{\dagger} +C.C.) 
+\frac{\bar\theta}{8m\varpi}\epsilon_{jm}\dot{h}_{jk}(a_m a_k +a_m a_{k}^{\dagger} +C.C.).
\label{e16}
\end{eqnarray}
where C.C. means complex conjugate. Hence the Heisenberg  eqn. of motion of $a_{j}(t)$ (\ref{evolution}) reads
\begin{eqnarray}
\frac{da_{j}(t)}{dt} &=& \frac {-i\varpi}{2}\left(a_j-a_j^\dagger\right) + 
\frac{1}{2}\dot h_{jk}a^\dagger_k + 
\frac{\bar{\theta}}{4m\hbar}\epsilon_{jk} a_{k}+\frac{im\varpi\theta}{8\hbar}
(\epsilon_{lj}\dot h_{lk}+\epsilon_{lk}\dot h_{lj})(a_k - a_{k}^{\dagger})\nonumber\\
&&-\frac{i\bar{\theta}}{8m\varpi\hbar}
(\epsilon_{lj}\dot h_{lk}+\epsilon_{lk}\dot h_{lj})(a_k + a_{k}^{\dagger})
\label{e17fp}
\end{eqnarray}
and that of $a_{j}^{\dagger}(t)$, given by the C.C of the above equation.
Using eq(s)( \ref{e19}) the time-evolution described in (\ref{e17fp}) can be cast  
in terms of the matrix pair $\left( \zeta, \xi \right) $ defined in (\ref{matrix_pair}):
\begin{eqnarray}
\frac{d\zeta_{jk}}{dt}=
-\frac{1}{2}{\dot h}_{jl}\zeta_{lk} + \frac {\bar{\theta}}{4m\hbar}\epsilon_{jl}\zeta_{lk}
- \frac{i\bar{\theta}}{4m\varpi\hbar}
(\epsilon_{lj}\dot h_{lp}+\epsilon_{lp}\dot h_{lj})\xi_{pk}\>
\label{e21a}\\
\frac{d \xi_{jk}}{dt} = -i\varpi \zeta_{jk} + 
\frac{1}{2}{\dot h}_{jl}\xi_{lk} 
+\frac{im\varpi\theta}{4\hbar}(\epsilon_{pl}\dot h_{jp}-\epsilon_{jp}\dot h_{pl})\zeta_{lk}
+\frac{\bar{\theta}}{4m\hbar}\epsilon_{jl}\xi_{lk}~.
\label{e21b} 
\end{eqnarray}
Since we want to consider interacting with circularly polarized GW, we substitute eqns.$(\ref{e13},\ref{e23},\ref{e33},\ref{e43})$ in eqns.$(\ref{e21a},\ref{e21b})$ and get the quantum dynamics of the system described in terms of a set of first order differential eqns 
\begin{eqnarray}
\frac{dA}{dt} + f_0\Omega C -\Lambda_{\bar{\theta}}D +\frac{i4 \Lambda_{\bar{\theta}}}{  \varpi}f_0\Omega F &=& 0\>,
\nonumber \\
\frac{dB}{dt} - \Omega C-f_0\Omega D+\Lambda_{\bar{\theta}}C +\frac{i4 \Lambda_{\bar{\theta}}}{  \varpi}f_0\Omega E&=& 0\>,
\nonumber \\
\frac{dC}{dt} + \Omega B + f_0\Omega A-\Lambda_{\bar{\theta}}B+\frac{i4 \Lambda_{\bar{\theta}}}{  \varpi}f_0\Omega H &=& 0\>,
\nonumber \\
\frac{dD}{dt} - f_0\Omega B+\Lambda_{\bar{\theta}}A+\frac{i4 \Lambda_{\bar{\theta}}}{  \varpi}f_0\Omega G &=& 0\>
\nonumber \\
\frac{dE}{dt} +i\varpi A - f_0\Omega G-\Lambda_{\bar{\theta}}H -i\Lambda f_0\Omega B  &=& 0\>,
\nonumber \\
\frac{dF}{dt} - \Omega G+i\varpi B + f_0\Omega H+\Lambda_{\bar{\theta}}G -i\Lambda f_0\Omega A  &=& 0\>,
\nonumber \\
\frac{dG}{dt} + \Omega F +i\varpi C - f_0\Omega E-\Lambda_{\bar{\theta}}F -i\Lambda f_0\Omega D &=& 0\>,
\nonumber \\
\frac{dH}{dt} +i\varpi D + f_0\Omega F+\Lambda_{\bar{\theta}}E -i\Lambda f_0\Omega C  &=& 0\>.
\label{e53}
\end{eqnarray}
to be solved to first order in the GW amplitude with boundary conditions eq$(\ref{bc0})$, which physically signifies that the GW hits the particle at t=0. 
Here 
\begin{eqnarray}
\Lambda_{\bar{\theta}} = \frac{\bar{\theta}}{4m\hbar}
\label{Lambda_1}
\end{eqnarray}
is a new frequency that is characteristic of the NC momentum scale $\bar{\theta}$ and 
\begin{eqnarray}
\Lambda = \frac{m\varpi \theta}{\hbar}
\label{Lambda_2}
\end{eqnarray}
is a dimensionless parameter with spatial NC parameter $\sqrt{\theta}$.

Solving the eq.(s)$(\ref{e53})$ we get 
\begin{eqnarray}
A(t) &=& \cos\Lambda_{\bar{\theta}}t -f_0\Omega\frac{1}{\Omega-\Lambda_{\bar{\theta}}}\left(N_3+\frac{\varpi}{\Omega-\Lambda_{\bar{\theta}}}\right)-4\Lambda_{\bar{\theta}}f_0\Omega\frac{N_2}{\varpi\left(\Omega-\Lambda_{\bar{\theta}}\right)}-i\frac{4\Lambda_{\bar{\theta}}f_0\Omega N_1}{\varpi\left(\Omega-\Lambda_{\bar{\theta}}\right)}\nonumber\\
B(t) &=&N_1-f_0\Omega\frac{\left(1-\cos\Lambda_{\bar{\theta}}t\right)}{\Lambda_{\bar{\theta}}}-4f_0\Omega\left[t\sin\Lambda_{\bar{\theta}}t+\frac{1-\cos\Lambda_{\bar{\theta}}t}{\Lambda_{\bar{\theta}}}\right]-i\frac{4f_0\Omega\sin\Lambda_{\bar{\theta}}t}{\varpi}\nonumber\\
C(t) &=&-N_1-\frac{f_0\Omega\sin\Lambda_{\bar{\theta}}t}{\Lambda_{\bar{\theta}}}-4f_0\Omega\left[t\sin\Lambda_{\bar{\theta}}t+\frac{1-\cos\Lambda_{\bar{\theta}}t}{\Lambda_{\bar{\theta}}}\right]-i\frac{4f_0\Omega\sin\Lambda_{\bar{\theta}}t}{\varpi}\nonumber\\
D(t) &=&-\sin\Lambda_{\bar{\theta}}t+f_0\Omega\frac{N_1}{\Omega-\Lambda_{\bar{\theta}}}-4f_0\Omega \frac{\Lambda_{\bar{\theta}}}{\varpi\left(\Omega-\Lambda_{\bar{\theta}}\right)}\left(N_2-\frac{\varpi}{\Omega-\Lambda_{\bar{\theta}}}\right)-i\frac{4\Lambda_{\bar{\theta}}f_0\Omega N_1}{\varpi\left(\Omega-\Lambda_{\bar{\theta}}\right)}\nonumber\\
E(t) &=& \cos\Lambda_{\bar{\theta}}t+f_0\Omega \frac{N_1}{\Omega-\Lambda_{\bar{\theta}}}+i\left[ -\frac{\varpi \sin\Lambda_{\bar{\theta}} t}{\Lambda_{\bar{\theta}}}+\frac{f_0 \varpi}{\Omega-\Lambda_{\bar{\theta}}}\left(N_2 - \frac{\varpi}{\Omega-\Lambda_{\bar{\theta}}}\right)-\varpi\left(t\cos\Lambda_{\bar{\theta}} t +\frac{\sin\Lambda_{\bar{\theta}} t}{\Lambda_{\bar{\theta}}}\right)\right.\nonumber\\
&&\left. +\Lambda f_0 \Omega \frac{N_1}{\Omega-\Lambda_{\bar{\theta}}}\right]\nonumber\\
F(t) &=&N_1- f_0\Omega\frac{\left(1-\cos\Lambda_{\bar{\theta}}t\right)}{\Lambda_{\bar{\theta}}}+i\left[\left(N_2-\frac{\varpi}{\Omega-\Lambda_{\bar{\theta}}}\right)-\frac{\varpi N_1}{\Omega-\Lambda_{\bar{\theta}}}+\Lambda f_0\Omega\frac{\sin\Lambda_{\bar{\theta}} t}{\Lambda_{\bar{\theta}}}-f_0\Omega\varpi t \frac{\sin\Lambda_{\bar{\theta}} t}{\Lambda_{\bar{\theta}}}\right.\nonumber\\
&&\left. +f_0\Omega \varpi\frac{\left(1-\cos\Lambda_{\bar{\theta}} t\right)}{\left(\Lambda_{\bar{\theta}}\right)^2} \right]\nonumber\\
G(t) &=&-N_1+f_0\Omega\frac{\sin\Lambda_{\bar{\theta}}t}{\Lambda_{\bar{\theta}}}+i\left[\left(N_3+\frac{\varpi}{\Omega-\Lambda_{\bar{\theta}}}\right)-\frac{\varpi N_1}{\Omega-\Lambda_{\bar{\theta}}}+\Lambda f_0\Omega\frac{1-\cos\Lambda_{\bar{\theta}}t}{\Lambda_{\bar{\theta}}}-\varpi t f_0\Omega\frac{\sin\Lambda_{\bar{\theta}}t}{\Lambda_{\bar{\theta}}}\right.\nonumber\\
&&\left. -f_0\Omega\varpi\frac{1-\cos\Lambda_{\bar{\theta}}t}{\Lambda_{\bar{\theta}}^2}\right]\nonumber\\
H(t) &=&-\sin\Lambda_{\bar{\theta}} t -\frac{f_0\Omega N_1}{\Omega-\Lambda_{\bar{\theta}}}+i\left[\frac{f_0\Omega}{\Omega-\Lambda_{\bar{\theta}}}\left(N_3+\frac{\varpi}{\Omega-\Lambda_{\bar{\theta}}}\right)-\varpi t \sin\Lambda_{\bar{\theta}} t +\frac{\varpi\left(1-\cos\Lambda_{\bar{\theta}} t\right)}{\Lambda_{\bar{\theta}}}\right.\nonumber\\
&&\left. +\Lambda f_0\Omega\frac{N_1}{\Omega-\Lambda_{\bar{\theta}}}\right]
 \label{100z}
\end{eqnarray}
where $N_1$, $N_2$, $N_3$ are given by
\begin{eqnarray}
N_1 &=& 1-\cos\left(\Omega-\Lambda_{\bar{\theta}}\right)t\nonumber\\
N_2 &=& \varpi \left[\frac{\sin\left(\Omega-\Lambda_{\bar{\theta}}\right)t+\cos\left(\Omega-\Lambda_{\bar{\theta}}\right)t-1}{\Omega-\Lambda_{\bar{\theta}}}-t\right]\nonumber\\
N_3 &=& \varpi\left[\frac{\sin\left(\Omega-\Lambda_{\bar{\theta}}\right)t-\cos\left(\Omega-\Lambda_{\bar{\theta}}\right)t+1}{\Omega-\Lambda_{\bar{\theta}}}-t\right]\nonumber\\
\label{110}
\end{eqnarray}
Using these expressions in (\ref{e33}, \ref{e43}) we can have the solution in terms of the matrix pair $\left( \zeta, \xi \right)$ which can be further substituted in the in ( \ref{e19}) via (\ref{matrix_pair}) to obtain the time-evolution of the raising and lowering operators $a_{j} \left(t \right)$ and $ a_{j}^\dagger \left(t \right)$ in terms of their initial values $a_{j} \left(0 \right)$ and $a_{j}^\dagger \left(0\right)$.
Using the definition of the raising and lowering operators (\ref{xp}) at initial time and any subsequent time yields the expectation value of the components of position and momentum of the particle at an arbitrary time $t$ in terms of their initial expectation values $\left(X_{1}\left( 0 \right), X_{2}\left( 0 \right)\right)$ and momentum $\left(P_{1}\left( 0 \right), P_{2}\left( 0 \right)\right)$. We give the explicit expression for $\langle X_{1}\left(t\right)\rangle$ and $\langle X_{2}\left(t\right)\rangle$ of the test bodies below. 
\begin{eqnarray}
\langle X_{1}\left(t\right)\rangle & = &\left[\cos\Lambda_{\bar{\theta}} t+ \left(\epsilon_3-\epsilon_1\right)N_1\right]X_{1}\left( 0 \right)+\left[\sin\Lambda_{\bar{\theta}}t+ \left(\epsilon_1+\epsilon_3\right)N_1\right]X_{2}\left( 0 \right) \nonumber\\
&+&\left[\varpi t \cos\Lambda_{\bar{\theta}}t + \left(N_2\epsilon_3+N_3\epsilon_3\right) \right]\frac{P_{1}\left( 0 \right)}{m\varpi}\nonumber\\
&&+\left[-\varpi\left(\Omega-\Lambda_{\bar{\theta}}\right)\left.\{\frac{t\left(1-\cos\Lambda_{\bar{\theta}} t\right)}{\Lambda_{\bar{\theta}}}-\frac{\sin\Lambda_{\bar{\theta}} t}{\Lambda_{\bar{\theta}}^2}\right.\}\epsilon_1-\left(N_2\epsilon_1+N_3\epsilon_3\right)
+\frac{\varpi}{\Lambda_{\bar{\theta}}}\left(1-\cos\Lambda_{\bar{\theta}} t\right)\epsilon_1\right]\frac{P_{2}\left( 0 \right)}{m\varpi}\nonumber\\
&& +f_0\left[-\frac{4\Lambda_{\bar{\theta}}\Omega}{\varpi\left(\Omega-\Lambda_{\bar{\theta}}\right)}\left(N_3-\frac{\varpi}{\Omega-\Lambda_{\bar{\theta}}}\right)-\frac{\Omega N_1}{\Omega-\Lambda_{\bar{\theta}}} -4\Omega\left(\epsilon_1+\epsilon_3\right)\left.\{t\sin\Lambda_{\bar{\theta}} t -\frac{1-\cos\Lambda_{\bar{\theta}} t}{\Lambda_{\bar{\theta}}}\right.\}\right.\nonumber\\
&& \, \, \, \, \, \left. -\Omega t \epsilon_1 \cos\Lambda_{\bar{\theta}}t \right]X_{1}\left( 0 \right)\nonumber\\
&&+f_0\left[-4\Omega t \sin\Lambda_{\bar{\theta}} t\left(\epsilon_1-\epsilon_3\right)-\frac{\Omega N_1}{\Omega-\Lambda_{\bar{\theta}}}+\frac{4 \Lambda_{\bar{\theta}} \Omega}{\varpi \left(\Omega-\Lambda_{\bar{\theta}}\right)}\left(N_2-\frac{\varpi}{\Omega-\Lambda_{\bar{\theta}}}\right) -\Omega t \epsilon_3\cos\Lambda_{\bar{\theta}}t \right]X_{2}\left( 0 \right)\nonumber\\
&&+f_0\left[-\frac{\Omega}{\Omega-\Lambda_{\bar{\theta}}}\left(N_2-\frac{\varpi}{\Omega-\Lambda_{\bar{\theta}}}\right)+\frac{4\Lambda_{\bar{\theta}}\Omega N_1}{\varpi\left(\Omega-\Lambda_{\bar{\theta}}\right)}+\frac{\Lambda \Omega}{\Lambda_{\bar{\theta}}}\left.\{\epsilon_3\sin\Lambda_{\bar{\theta}} t +\epsilon_1\left(1-\cos\Lambda_{\bar{\theta}} t\right)\right.\}
\right.\nonumber\\
&&\left.-\frac{\Lambda\Omega N_1}{\Omega-\Lambda_{\bar{\theta}}}-\frac{\Omega \varpi t}{\Lambda_{\bar{\theta}}}\left.\{\epsilon_3\left(1-\cos\Lambda_{\bar{\theta}}t\right)-\epsilon_1\sin\Lambda_{\bar{\theta}}t\right.\}+\frac{\Omega \varpi}{\left(\Lambda_{\bar{\theta}}\right)^2}\left.\{-\epsilon_3\sin\Lambda_{\bar{\theta}} t+\epsilon_1\left(1-\cos\Lambda_{\bar{\theta}} t\right)\right.\}\right.\nonumber\\
&&\left.-\Lambda\Omega t \epsilon_3 \cos\Lambda_{\bar{\theta}}t + \frac{\varpi\Omega t^2\epsilon_1}{2}\cos\Lambda_{\bar{\theta}} t -\frac{4\Omega\sin\Lambda_{\bar{\theta}} t}{\varpi}\left(\epsilon_3+\epsilon_1\right)\right]\frac{P_{1}\left( 0 \right)}{m\varpi}\nonumber\\
&&+f_0\left[-\frac{\Lambda\Omega}{\Lambda_{\bar{\theta}}}\left.\{\epsilon_1\sin\Lambda_{\bar{\theta}} t-\epsilon_3\left(1-\cos\Lambda_{\bar{\theta}}t\right)\right.\}-\frac{\Lambda_{\bar{\theta}}\Omega N_1}{\Omega-\Lambda_{\bar{\theta}}}\epsilon_1+\frac{\Omega\varpi t}{\Lambda_{\bar{\theta}}}\left.\{\epsilon_1\left(1-\cos\Lambda_{\bar{\theta}}t\right)+\epsilon_3\sin\Lambda_{\bar{\theta}} t\right.\}\right.\nonumber\\
&&\left.+\frac{\Lambda\Omega}{\left(\Lambda_{\bar{\theta}}\right)^2}\left.\{-\epsilon_1\sin\Lambda_{\bar{\theta}}t+\epsilon_3\left(1-\cos\Lambda_{\bar{\theta}}t\right)\right.\}+\frac{4\Omega\sin\Lambda_{\bar{\theta}}t}{\varpi}\left(\epsilon_1-\epsilon_3\right)-\frac{\Omega}{\Omega-\Lambda_{\bar{\theta}}}\left(N_3+\frac{\varpi}{\Omega-\Lambda_{\bar{\theta}}}\right)\right.\nonumber\\
&&\left.-\frac{\Lambda\Omega N_1}{\Omega-\Lambda_{\bar{\theta}}}-\frac{4\Lambda_{\bar{\theta}}\Omega N_1}{\varpi\left(\Omega-\Lambda_{\bar{\theta}}\right)}-\Lambda\Omega t \epsilon_1\cos\Lambda_{\bar{\theta}}t +\frac{\varpi\Omega t ^2\epsilon_3}{2}\cos\Lambda_{\bar{\theta}}t\right]\frac{P_{2}\left( 0 \right)}{m\varpi}
\label{fpx1}
\end{eqnarray}
\begin{eqnarray}
\langle X_{2}\left(t\right)\rangle & = &\left[\cos\Lambda_{\bar{\theta}} t- \left(\epsilon_3-\epsilon_1\right)N_1\right]X_{2}\left( 0 \right)+\left[\sin\Lambda_{\bar{\theta}} t+ \left(\epsilon_1+\epsilon_3\right)N_1\right]X_{1}\left( 0 \right)\nonumber\\
&&+\left[ \varpi t \cos\Lambda_{\bar{\theta}}t + \left(N_2\epsilon_3-N_3\epsilon_3\right) \right] \frac{P_{2}\left( 0 \right)}{m\varpi}\nonumber\\
&&+\left[-\varpi\left(\Omega-\Lambda_{\bar{\theta}}\right)\left.\{\frac{t\left(1-\cos\Lambda_{\bar{\theta}} t\right)}{\Lambda_{\bar{\theta}}} - \frac{\sin\Lambda_{\bar{\theta}} t}{\Lambda_{\bar{\theta}}^2}\right.\}\epsilon_1- \left(N_2\epsilon_1+N_3\epsilon_3\right)
+\frac{\varpi}{\Lambda_{\bar{\theta}}}\left(1-\cos\Lambda_{\bar{\theta}} t\right)\epsilon_1\right]\frac{P_{1}\left( 0 \right)}{m\varpi}\nonumber\\
&&+f_0\left[-\frac{4\Lambda_{\bar{\theta}}\Omega}{\varpi\left(\Omega-\Lambda_{\bar{\theta}}\right)}\left(N_3-\frac{\varpi}{\Omega-\Lambda_{\bar{\theta}}}\right)-\frac{\Omega N_1}{\Omega-\Lambda_{\bar{\theta}}} -4\Omega\left(\epsilon_1+\epsilon_3\right)\left.\{t\sin\Lambda_{\bar{\theta}} t -\frac{1-\cos\Lambda_{\bar{\theta}} t}{\Lambda_{\bar{\theta}}}\right.\} \right.\nonumber\\
&&\left.+\Omega t \epsilon_1 \cos\Lambda_{\bar{\theta}}t \right]X_{2}\left( 0 \right)\nonumber\\
&&+f_0\left[-4\Omega t \sin\Lambda_{\bar{\theta}} t\left(\epsilon_1-\epsilon_3\right)+\frac{\Omega N_1}{\Omega-\Lambda_{\bar{\theta}}}+\frac{4 \Lambda_{\bar{\theta}} \Omega}{\varpi \left(\Omega-\Lambda_{\bar{\theta}}\right)}\left(N_2-\frac{\varpi}{\Omega-\Lambda_{\bar{\theta}}}\right) +\Omega t \epsilon_3\cos\Lambda_{\bar{\theta}} t \right]X_{1}\left( 0 \right)\nonumber\\
&&+f_0\left[-\frac{\Omega}{\Omega-\Lambda_{\bar{\theta}}}\left(N_2-\frac{\varpi}{\Omega-\Lambda_{\bar{\theta}}}\right)+\frac{4\Lambda_{\bar{\theta}}\Omega N_1}{\varpi\left(\Omega-\Lambda_{\bar{\theta}}\right)}+\frac{\Lambda \Omega}{\Lambda_{\bar{\theta}}}\left.\{\epsilon_3\sin\Lambda_{\bar{\theta}} t +\epsilon_1\left(1-\cos\Lambda_{\bar{\theta}} t\right)\right.\}\right.\nonumber\\
&&\left.-\frac{\Lambda\Omega N_1}{\Omega-\Lambda_{\bar{\theta}}}-\frac{\Omega \varpi t}{\Lambda_{\bar{\theta}}}\left.\{\epsilon_3\left(1-\cos\Lambda_{\bar{\theta}}t\right)-\epsilon_1\sin\Lambda_{\bar{\theta}}t\right.\}+\frac{\Omega \varpi}{\left(\Lambda_{\bar{\theta}}\right)^2}\left.\{-\epsilon_3\sin\Lambda_{\bar{\theta}} t+\epsilon_1\left(1-\cos\Lambda_{\bar{\theta}} t\right)\right.\}\right.\nonumber\\
&&\left.+\Lambda\Omega t \epsilon_3 \cos\Lambda_{\bar{\theta}}t + \frac{\varpi\Omega t^2\epsilon_1}{2}\cos\Lambda_{\bar{\theta}} t -\frac{4\Omega\sin\Lambda_{\bar{\theta}} t}{\varpi}\left(\epsilon_3+\epsilon_1\right)\right]\frac{P_{2}\left( 0 \right)}{m\varpi}\nonumber\\
&&+f_0\left[-\frac{\Lambda\Omega}{\Lambda_{\bar{\theta}}}\left.\{\epsilon_1\sin\Lambda_{\bar{\theta}} t-\epsilon_3\left(1-\cos\Lambda_{\bar{\theta}}t\right)\right.\}-\frac{\Lambda_{\bar{\theta}}\Omega N_1}{\Omega-\Lambda_{\bar{\theta}}}\epsilon_1+\frac{\Omega\varpi t}{\Lambda_{\bar{\theta}}}\left.\{\epsilon_1\left(1-\cos\Lambda_{\bar{\theta}}t\right)+\epsilon_3\sin\Lambda_{\bar{\theta}} t\right.\}\right.\nonumber\\
&&\left.+\frac{\Lambda\Omega}{\left(\Lambda_{\bar{\theta}}\right)^2}\left.\{-\epsilon_1\sin\Lambda_{\bar{\theta}}t+\epsilon_3\left(1-\cos\Lambda_{\bar{\theta}}t\right)\right.\}+\frac{4\Omega\sin\Lambda_{\bar{\theta}}t}{\varpi}\left(\epsilon_1-\epsilon_3\right)+\frac{\Omega}{\Omega-\Lambda_{\bar{\theta}}}\left(N_3+\frac{\varpi}{\Omega-\Lambda_{\bar{\theta}}}\right)\right.\nonumber\\
&&\left.-\frac{\Lambda\Omega N_1}{\Omega-\Lambda_{\bar{\theta}}}-\frac{4\Lambda_{\bar{\theta}}\Omega N_1}{\varpi\left(\Omega-\Lambda_{\bar{\theta}}\right)}-\Lambda\Omega t \epsilon_1\cos\Lambda_{\bar{\theta}}t -\frac{\varpi\Omega t ^2\epsilon_3}{2}\cos\Lambda_{\bar{\theta}}t\right]\frac{P_{1}\left( 0 \right)}{m\varpi}
\label{fpx2}
\end{eqnarray}
Let us try to point out the salient features of this result. First of all notice that there is a smooth commutative limit $\left( \theta, \bar{\theta}\right) \to \left(0, 0 \right)$ i.e., $\left( \Lambda, \Lambda_{\bar{\theta}} \right) \to \left(0, 0 \right)$ to these solutions owing to the factors like $\frac{\left(1-\cos\Lambda_{\bar{\theta}} t\right)}{\Lambda_{\bar{\theta}}}$ and $\frac{\sin \Lambda_{\bar{\theta}} t}{\Lambda_{\bar{\theta}}}$.

Next, we look at the different terms of the above solutions. The first four square bracketed terms in both eqn.s (\ref{fpx1}, \ref{fpx2}) show that the momentum noncommutativity makes the solution oscillatory with a characteristic frequency $\Lambda_{\bar{\theta}}$ even in the absence of any GW. 
The terms proportional to $N_{1}, N_{2}$  and$N_{3}$ appear there since we have chosen a directional triad which is rotating with a frequency $\Omega$. Further, for a non-zero $\bar{\theta}$ the free particle sees the triad to rotate with a reduced frequency $\left(\Omega-\Lambda_{\bar{\theta}}\right)$ instead of $\Omega$. 

Rest of the terms are proportional to the GW amplitude $f_{0}$ and show the manifest effect of GW. In both eqn. (\ref{fpx1}) and (\ref{fpx2}), among the terms with a $\left(\Omega-\Lambda_{\bar{\theta}}\right)^{-1}$ factor, some will grow rapidly near $\Omega = \Lambda_{\bar{\theta}}$ for non-zero $\bar{\theta}$. This we refer to as the resonance behavior. Note that the resonance appears here typically between the oscillatory nature induced into the free-particle motion by the noncommutativity among the momenta and the periodic GW. Thus it is a noncommutative effect. 

Such noncommutativity-induced resonance behaviour appeared earlier in \cite{ncgw3} (where we considered response of a free particle to linearly polarized GW) at the same value of frequency $\Lambda_{\bar{\theta}}=\frac{\bar{\theta}}{4m\hbar} = \left( \frac{m}{m_{n}}\right)^{-1}0.33 Hz$. This sub-Hz frequency range is estimated using the latest upper bound estimation of the momentum NC parameter \cite{bert0} $\bar{\theta} \leq 2.32 \times 10^{-61} {\rm kg^{2}m^{2}sec^{-2}}$ and taking neutron mass $m_{n}= 167.32 \times 10^{-29}$kg) as the reference mass of a free particle. Reason for choosing neutron mass as reference is that we are considering the quantum dynamics of a free particle and only microscopic particles qualify. Thus we see from the present paper that the resonance point does not depend on the kind of polarization of the GW. This answers the first question posed in the introduction at least for the free particle case.

Also note that the resonance is likely to occur in the milli-Hz range. Therefore, the present ground-based detectors operating in the kHz range will not be sensitive to them.
However,  the upcoming space-based GW detectors (e.g. LISA \cite{lisa}) will look precisely in the milli-Hz frequency range and should be able to pick up such NC effects. Also the recently proposed Torsion-Bar Antenna (TOBA) \cite{TOBA1, TOBA2} for low-frequency GW observations may be relevant in this context.
Recently the combined atom and laser interferometry technique has been proposed that also claims to extend the GW detection frequency band to the lower frequency region namely 0.1-10 Hz. 
  
We finish this section by noticing that whereas the characteristic frequency $\Lambda_{\bar{\theta}}$ due to momentum noncommutativity is ubiquitous in both eqn.s (\ref{fpx1}, \ref{fpx2}), the dimensionless parameter $\Lambda$ due to coordinate noncommutativity only appears in a handful of terms, and such terms always carry the amplitude of the GW $f_{0}$. Using the upper-bound estimate\cite{carol} of the spatial NC parameter $\theta \approx 10^{-40} {\rm m^{2}}$ and neutron mass as reference, this parameter value is estimated as $\Lambda = \frac{m\varpi \theta}{\hbar} \approx \varpi \left(\frac{m}{m_{n}}\right)10^{-33}$. Since $\varpi$ is determined from the initial uncertainty of the momentum or position of the quantum particle, it can not be large enough to make $\Lambda$ appreciable. Further, the smallness of GW strain amplitude $f_{0} \approx 10^{-23}$ supresses such terms even further. 
That spatial noncommutativity does not affect the response of a free particle to GW in a significant manner, a result established earlier in \cite{ncgw1, ncgw3} for linearly polarized GWs, is thus shown to hold for circularly polarized GWs as well, in the present paper, again answering the first question raised in the introduction.

In the next section we will consider the response of the more non-trivial HO system to both linear and circularly polarized GW in a NC phase-space.
\section{Harmonic oscillator in NC phase space interacting with linearly and circularly polarized gravitational wave}
For the harmonic oscillator case the potential function is $V = \frac{1}{2}m \varpi^{2} x^{2}$ and the Hamiltonian, in terms of the raising and lowering operators, takes the form 
\begin{eqnarray}
{\hat H} &=& \hbar\varpi\left( a_j^\dagger a_j + 1 \right) 
- \frac{i\hbar}{4} \dot h_{jk} 
\left(a_j a_k - a_j^\dagger a_k^\dagger\right)  + \frac{m \varpi \theta}{8} \epsilon_{jm} {\dot h}_{jk}  \left(a_{m}a_{k}  - a_{m}a_{k}^\dagger + C.C \right)  \nonumber \\ 
&& -\frac{i}{2} m \varpi ^2 \theta\epsilon_{jk}a_{j}^\dagger a_{k}+ \frac {i\bar{\theta}}{4m}\epsilon_{jk}a_j^\dagger a_k+\frac{\bar\theta}{8m\varpi}\epsilon_{jm}\dot{h}_{jk}(a_m a_k +a_m a_{k}^{\dagger} +C.C.).\>
\label{e16}
\end{eqnarray}
where C.C means complex conjugate as usual. Again working in the Heisenberg representation, the time evolution of $a_{j}(t)$ is given by 
\begin{eqnarray}
\frac{da_{j}(t)}{dt} &=& -i{\varpi}a^j + 
\frac{1}{2}\dot h_{jk}a^\dagger_k - 
\frac{m\varpi^{2}\theta}{2\hbar}\epsilon_{jk} a_{k}+ \frac{i m \varpi \theta}{8 \hbar} \left(\epsilon_{lj} {\dot h}_{lk} 
+ \epsilon_{lk} {\dot h}_{lj}\right)\left(a_{k} - a^\dagger_{k}\right)\nonumber \\
&& + \frac{\bar{\theta}}{4m\hbar}\epsilon_{jk} a_{k}-\frac{i\bar{\theta}}{8m\varpi\hbar}
(\epsilon_{lj}\dot h_{lk}+\epsilon_{lk}\dot h_{lj})(a_k + a_{k}^{\dagger})\>
\label{e17}
\end{eqnarray}
and that of $a_{j}^{\dagger}(t)$ is the C.C of the above equation. In terms of the matrix pair $\left( \zeta, \xi \right)$ in  (\ref{matrix_pair}) the time-evolution equations take the form
\begin{eqnarray}
\frac{d \zeta_{jk}}{dt}&=&-i\varpi \xi_{jk} 
-\frac{1}{2}{\dot h}_{jl}\zeta_{lk} - 
\frac{m \varpi^{2}\theta}{2 \hbar}\epsilon_{jl} \zeta_{lk} + \frac {\bar{\theta}}{4m\hbar}\epsilon_{jl}\zeta_{lk}
- \frac{i\bar{\theta}}{4m\varpi\hbar}
(\epsilon_{lj}\dot h_{lp}+\epsilon_{lp}\dot h_{lj})\xi_{pk}\> 
\label{e21aa}\\
\frac{d \xi_{jk}}{dt}& = &-i\varpi \zeta_{jk} + 
\frac{1}{2}{\dot h}_{jl}\xi_{lk} +
\Theta_{jl} \zeta_{lk} - \frac{m \varpi^{2}\theta}{2 \hbar}
\epsilon_{jl} \xi_{lk}+\frac{\bar{\theta}}{4m\hbar}\epsilon_{jl}\xi_{lk}\> 
\label{e21bb} 
\end{eqnarray}
where $\Theta_{jl}$ is the term reflecting 
the interplay of noncommutativity with GW
\begin{eqnarray}\Theta_{jl} = 
\frac{i m \varpi \theta}{4 \hbar}\left({\dot h}_{jm}
\epsilon_{ml} - \epsilon_{jm} {\dot h}_{ml}\right)\>.
\label{e21ab}
\end{eqnarray}
In the following we shall solve eq(s) $(\ref{e21aa}, \ref{e21bb})$ for the special cases of linearly  and circularly polarized GW respectively. 
\subsection{Response to linearly polarized GW}
Referring to (\ref{e13a}) and (\ref{form2}, \ref{form1}) we substitute for $h_{ij}$ and $\left( \zeta, \xi \right)$
in eq(s) $(\ref{e21aa}, \ref{e21bb})$ and comparing the coefficients of 
$I$ and $\sigma$-matrices, we get a set of 
first order differential equations for 
$A, B_{1}, B_{2}, B_{3}, C, D_{1}, D_{2}, D_{3}$ :
\begin{eqnarray}
\dot{A} &=& - i \varpi C - \dot{f}\left(\varepsilon_{1}B_{1} 
+ \varepsilon_{3}B_{3}\right)-4i\frac{\Lambda_{\bar{\theta}}}{\varpi} \dot{f}(\varepsilon_{3}D_{1}-\varepsilon_{1}D_{3}) - i \Lambda_0 B_{2} \nonumber\\
\dot{B}_{1} &=& - i \varpi D_{1} - \dot{f}\left(\varepsilon_{1} A -
 i\varepsilon_{3}B_{2}\right)-4\frac{\Lambda_{\bar{\theta}}}{\varpi} \dot{f}(i\varepsilon_{3}C-\varepsilon_{1}D_{2}) +  \Lambda_0 B_{3} \nonumber\\
\dot{B}_{2}  &=& - i \varpi D_{2} - i \dot{f}\left(\varepsilon_{3}B_{1} - 
\varepsilon_{1}B_{3}\right) -4\frac{\Lambda_{\bar{\theta}}}{\varpi} \dot{f}(\varepsilon_{1}D_{1}+\varepsilon_{3}D_{3})- i \Lambda_0 A \nonumber\\ 
\dot{B}_{3}&=& - i \varpi D_{3} - \dot{f}\left(\varepsilon_{3} A 
+ i \varepsilon_{1}B_{2}\right) +4\frac{\Lambda_{\bar{\theta}}}{\varpi} \dot{f}(i\varepsilon_{1}C+\varepsilon_{3}D_{2}) - \Lambda_0 B_{1} \nonumber\\ 
\dot{C} &=& - i \varpi A + \dot{f}\left(\varepsilon_{1}D_{1} 
+ \varepsilon_{3}D_{3}\right) +  i \Lambda \dot{f} 
\left(\varepsilon_{3} B_{1} - \varepsilon_{1} B_{3}\right)-i \Lambda_0 D_{2} \nonumber \\ 
\dot{D}_{1} &=& - i \varpi B_{1} + 
\dot{f}\left(\varepsilon_{1}C - i \varepsilon_{3}D_{2}\right) 
+  \Lambda \dot{f} \left(i \varepsilon_{3} A -\varepsilon_{1} B_{2} \right)+ \Lambda_0 D_{3} \nonumber \\ 
\dot{D}_{2} &=& - i \varpi B_{2} + i \dot{f} \left(\varepsilon_{3}D_{1} 
- \varepsilon_{1}D_{3}\right) +  \Lambda \dot{f} 
\left( \varepsilon_{1} B_{1} + \varepsilon_{3} B_{3} \right) - i \Lambda_0 C \nonumber \\ 
\dot{D}_{3} &=& - i \varpi  B_{3} + 
\dot{f}\left(\varepsilon_{3} C + i \varepsilon_{1} D_{2}\right) 
-  \Lambda \dot{f} \left( i\varepsilon_{1} A + \varepsilon_{3} B_{2} \right) -  \Lambda_0 D_{1}  
\label{iteration8} 
\end{eqnarray}
Here dot represents derivative with respect to time $t$. $\Lambda$ is the same dimensionless parameter carrying the spatial NC scale defined in (\ref{Lambda_2}) and 
\begin{equation}
\Lambda_{0}=  \Lambda_{\theta}  - \Lambda_{\bar{\theta}}
\label{Lambda_zero}
\end{equation}
is the difference between the frequency characterized by the spatial NC scale $\theta$ given by 
\begin{eqnarray}
\Lambda_{\theta} = \frac{m \varpi^{2} \theta}{2 \hbar}
\label{Lambda}
\end{eqnarray}
and the frequency $\Lambda_{\bar{\theta}}$ characterized by the momentum NC scale $\bar{\theta}$ defined  earlier in (\ref{Lambda_1}). Once we arrive at the solutions to the present problem we shall provide estimates of these characteristic frequencies in context of the physical systems where the quantum harmonic oscillators we are discussing can be realized.

Noting that the GW strain amplitude is very small $|f(t)|<<1$, we again solve the above set of equations iteratively about its $f(t)=0$ solution. The appropriate boundary conditions to apply are again (\ref{bc0}, \ref{bc}). We obtain the following solutions to first order in the gravitational wave amplitude. We also restrict ourselves to first-order in the NC parameter:
\begin{eqnarray}
A(t) & = & C(t)= e^{-i\varpi t}\cos(\Lambda_0 t)
\label{001}\\
B_{2}(t) & = & D_{2}(t) = -ie^{-i\varpi t}\sin(\Lambda_0 t)
\label{002}\\
B_{1}(t)&=&\left(\epsilon_1+i \frac{4\Lambda_{\bar{\theta}}\epsilon_3}{\varpi}\right)K_1+\left(\epsilon_3+i \frac{4\Lambda_{\bar{\theta}}\epsilon_1}{\varpi}\right)K_2
\label{003}\\
B_{3}(t)&=&\left(\epsilon_3+i \frac{4\Lambda_{\bar{\theta}}\epsilon_1}{\varpi}\right)K_1-\left(\epsilon_1+i \frac{4\Lambda_{\bar{\theta}}\epsilon_3}{\varpi}\right)K_2
\label{004}\\
D_{1}(t)&=&-\left(\epsilon_1+\frac{i \Lambda\epsilon_3}{4}\right)K_1-\left(\epsilon_3+\frac{i \Lambda\epsilon_1}{4}\right)K_2
\label{005}\\
D_{3}(t)&=&-\left(\epsilon_3+\frac{i \Lambda\epsilon_1}{4}\right)K_1+ \left(\epsilon_1+\frac{i \Lambda\epsilon_3}{4}\right)K_2 \label{006}
\end{eqnarray}
with
\begin{eqnarray}
K_1 &=&-e^{-i\varpi t}\cos(\Lambda_0 t)f(t)-2i\varpi\int^{t}_{0}dt^{\prime}
e^{i\varpi (t - t^{\prime})}\cos(\Lambda_0 t^{\prime})f(t^{\prime})+ \varpi^{2} \int^{t}_{0} h_1 {(t^{\prime})}dt^{\prime}
\label{007}\\
K_2 &=&e^{-i\varpi t}\sin(\Lambda_0 t)f(t)+2i\varpi\int^{t}_{0}dt^{\prime}
e^{i\varpi (t - t^{\prime})}\sin(\Lambda_0 t^{\prime})f(t^{\prime})- \varpi^{2} \int^{t}_{0} h_2 {(t^{\prime})}dt^{\prime}
\label{008}
\end{eqnarray}
and
\begin{eqnarray}
h_{1}(t)=\int^{t}_{0} dt^{\prime}e^{-i\varpi t^{\prime}}
\cos(\Lambda_0 t^{\prime})f(t^{\prime})\nonumber\\
h_{2}(t)=\int^{t}_{0} dt^{\prime}e^{-i\varpi t^{\prime}}
\sin(\Lambda_0 t^{\prime})f(t^{\prime})~.
\label{007}
\end{eqnarray}
To work out the above integrals one needs specific GW wave form $f\left( t \right)$. In the present paper we have taken a monochromatic sinusoidal wave form representing periodic GW of frequency $\Omega$ 
\begin{equation}
f\left( t \right) = f_{0} e^{i \Omega t}
\label{linear_signal}
\end{equation}
Once we have the solutions (\ref{001} - \ref{006}) the system is essentially solved and all that remains is to compute all the way back to the expectation value of the components of position and momentum of the particle at an arbitrary time $t$ in terms of their initial expectation values $\left(X_{1}\left( 0 \right), X_{2}\left( 0 \right)\right)$ and momentum $\left(P_{1}\left( 0 \right), P_{2}\left( 0 \right)\right)$. How that can be done has been discussed in the last section (below (\ref{110})). We give the explicit expression for $\langle X_{1}\left(t\right)\rangle$ and $\langle X_{2}\left(t\right)\rangle$ of the harmonic oscillator mass here:
\begin{eqnarray}
 \langle X_{1}\left(t\right)\rangle & = & \frac{\left(\cos \varpi_{-} t + \cos \varpi_{+} t\right)}{2} X_{1}(0) + \frac{\left(\sin \varpi_{-} t + \sin \varpi_{+} t\right)}{2 m\varpi}P_{1}(0)\nonumber\\ 
&& + \frac{\left(\cos \varpi_{+} t - \cos \varpi_{-} t\right)}{2} X_{2}(0) + \frac{\left(\sin \varpi_{+} t - \sin \varpi_{-} t\right)}{2m\varpi}P_{2}(0) \nonumber\\
&&- f_0\left[q_1(t)-\frac{\Lambda}{4} q_4(t)\right]\left[\epsilon_3 X_{1}(0)+\epsilon_1 X_{2}(0)\right]+f_0\left[q_3(t)+\frac{\Lambda}{4} q_2(t)\right]\left[\epsilon_1 X_{1}(0)+\epsilon_3 X_{2}(0)\right] \nonumber\\
&&+f_0\left[-q_2(t)+\frac{4\Lambda_{\bar{\theta}}}{\varpi}q_3(t)\right]\frac{1}{m\varpi}\left[\epsilon_3 P_{1}(0)+\epsilon_1 P_{2}(0)\right]\nonumber\\
&& -f_0\left[q_4(t)+\frac{4\Lambda_{\bar{\theta}}}{\varpi}q_1(t)\right]\frac{1}{m\varpi}\left[\epsilon_1 P_{1}(0)-\epsilon_3 P_{2}(0)\right]
\label{x11}
\end{eqnarray}
\begin{eqnarray}
 \langle X_{2}\left(t\right)\rangle & = & \frac{\left(\cos \varpi_{-} t + \cos \varpi_{+} t\right)}{2} X_{2}(0) + \frac{\left(\sin \varpi_{-} t + \sin \varpi_{+} t\right)}{2 m\varpi}P_{2}(0)\nonumber\\ 
&& - \frac{\left(\cos \varpi_{+} t - \cos \varpi_{-} t\right)}{2} X_{1}(0) - \frac{\left(\sin \varpi_{+} t - \sin \varpi_{-} t\right)}{2m\varpi}P_{1}(0) \nonumber\\
&&+ f_0\left[q_1(t)-\frac{\Lambda}{4} q_4(t)\right]\left[\epsilon_3 X_{2}(0)-\epsilon_1 X_{1}(0)\right]+f_0\left[q_3(t)+\frac{\Lambda}{4} q_2(t)\right]\left[-\epsilon_1 X_{2}(0)+\epsilon_3 X_{1}(0)\right] \nonumber\\
&&+f_0\left[-q_2(t)+\frac{4\Lambda_{\bar{\theta}}}{\varpi}q_3(t)\right]\frac{1}{m\varpi}\left[-\epsilon_3 P_{2}(0)+\epsilon_1 P_{1}(0)\right]\nonumber\\
&& -f_0\left[q_4(t)+\frac{4\Lambda_{\bar{\theta}}}{\varpi}q_1(t)\right]\frac{1}{m\varpi}\left[-\epsilon_1 P_{2}(0)-\epsilon_3 P_{1}(0)\right]
\label{x22}
\end{eqnarray}
with the dimensionless functions 
\begin{eqnarray}
q_1\left(t\right) & = & -\frac{1}{2}\left(\cos \Delta\varpi_{+} t + \cos \Delta\varpi_{-} t\right)+\frac{\varpi}{\Delta\varpi^2-\Lambda_0^2}\left[-k_1(t)\cos\varpi t + k_2(t)\sin\varpi t\right]\nonumber\\
&& +\frac{\varpi^2}{2}\left[-\frac{\sin\Delta\varpi_-{t}}{\left(\Delta\varpi_-\right)^2}+\frac{\sin\Delta\varpi_+{t}}{\left(\Delta\varpi_+\right)^2}+\frac{4\Delta\varpi \Lambda_0}{\left(\Delta\varpi^2-\Lambda_0^2\right)^2}\right]
\label{q1}
\end{eqnarray}
\begin{eqnarray}
q_2\left(t\right) & = & -\frac{1}{2}\left(\sin \Delta\varpi_{+} t + \sin \Delta\varpi_{-} t\right)-\frac{\varpi}{\Delta\varpi^2-\Lambda_0^2}\left[k_1(t)\sin\varpi t + k_2(t)\cos\varpi t\right]\nonumber\\
&& +\frac{\varpi^2}{2}\left[\frac{\cos\Delta\varpi_-{t}}{\left(\Delta\varpi_-\right)^2}-\frac{\cos\Delta\varpi_+{t}}{\left(\Delta\varpi_+\right)^2}-\frac{2\Lambda_0 t}{\left(\Delta\varpi^2-\Lambda_0^2\right)}\right]
\label{q2}
\end{eqnarray}
\begin{eqnarray}
  q_3\left(t\right) & = & \frac{1}{2}\left(\sin \Delta\varpi_{+} t - \sin \Delta\varpi_{-} t\right)-\frac{\varpi}{\Delta\varpi^2-\Lambda_0^2}\left[k_1(t)\sin\varpi t - k_2(t)\cos\varpi t\right]\nonumber\\
&& +\frac{\varpi^2}{2}\left[-\frac{\sin\Delta\varpi_-{t}}{\left(\Delta\varpi_-\right)^2}-\frac{\sin\Delta\varpi_+{t}}{\left(\Delta\varpi_+\right)^2}+\frac{2\left(\Delta\varpi^2+\Lambda_0^2\right)}{\left(\Delta\varpi^2-\Lambda_0^2\right)^2}\right]
\label{q3}         
\end{eqnarray}
\begin{eqnarray}
 q_4\left(t\right) & = & -\frac{1}{2}\left(\cos \Delta\varpi_{+} t - \cos \Delta\varpi_{-} t\right)-\frac{\varpi}{\Delta\varpi^2-\Lambda_0^2}\left[k_1(t)\cos\varpi t + k_2(t)\sin\varpi t\right]\nonumber\\
&& -\frac{\varpi^2}{2}\left[\frac{\cos\Delta\varpi_-{t}}{\left(\Delta\varpi_-\right)^2}+\frac{\cos\Delta\varpi_+{t}}{\left(\Delta\varpi_+\right)^2}-\frac{2\Delta\varpi t}{\left(\Delta\varpi^2-\Lambda_0^2\right)}\right]
\label{q4}
\end{eqnarray}
and functions with the dimension of frequency
\begin{eqnarray}
k_1(t)&=&2\Delta\varpi-\Omega_-\cos\left(\Omega_+ t\right) +\Omega_+\cos\left(\Omega_- t\right) \label{k1}\\
k_2(t)&=&\Omega_+\sin\left(\Omega_- t \right)+\Omega_-\sin\left(\Omega_+ t\right) \label{k2}		
\end{eqnarray}
appear with the following frequencies
\begin{eqnarray}
\Delta\varpi &=& \varpi - \Omega \label{nat-gw} \, \left({\rm difference\, of\, the\, frequencies\, of\, the\, HO\, and \,that\, of\, the\, GW\, signal}\right)  \\
\varpi_{\pm}  &=& \varpi \pm \Lambda_0 \label{nat-pm-nc} \, \left({\rm natural\, frequency\, of\, the\, HO,\, shifted\, by\,} \Lambda_{0}\right)\\
\Omega_{\pm}  &=& \Omega \pm \Lambda_0 \label{gw-pm-nc}\, \left({\rm frequency\, of\, the\, GW,\, shifted\, by\,} \Lambda_{0}\right)\\
\Delta\varpi_{\pm}  &=& \Delta\varpi \pm \Lambda_0 \label{nat-gw-pm-nc}\, \left({\rm difference\, of\, the\, frequencies\,  of\, the\, HO\, and\, GW,\, shifted\, by\,} \Lambda_{0}\right)
\end{eqnarray}
Note that the the two frequencies $\Lambda_{\theta} $ and $\Lambda_{\bar{\theta}} $, characterizing respectively the noncommutativity of the spatial and the momentum sector, only appear as a combination $\Lambda_{0}$, defined in(\ref{Lambda_zero}) and hereafter referred as the characteristic frequency of the NC phase-space, in all sinusoidal terms 
As in the free particle case we now look at the different terms in the solutions (\ref{x11}, \ref{x22}) critically.

The first four terms in both (\ref{x11}, \ref{x22}) are independent of any GW effect and only show standard HO oscillator solution with the natural frequency $\varpi$ shifted by the characteristic frequency coming from NC phase-space. The remaining terms are the combined effect of the GW and the NC phase-space. Owing to the smallness of the GW strain amplitude $f_{0}$ these terms are ordinarily small. However, when the natural frequency of the HO becomes 
\begin{equation}
\varpi = \Omega_{\pm} = \Omega \pm \Lambda_0
\label{resonance}
\end{equation}
some terms in the dimensionless functions in (\ref{q1}, \ref{q2}, \ref{q3}, \ref{q4}), owing to the presence of factors $\Delta\varpi_{\pm} $ in the denominator, will show resonance; thus amplifying the effect of the GW. This result is consistent with our earlier result in \cite{ncgw2} where only spatial noncommutativity has been considered and the result there showed that due to the presence of spatial NC structure the ordinarily expected unique resonance point $\varpi = \Omega$ will split into two evenly spaced resonance points $\varpi = \Omega \pm \Lambda_{\theta}$. Here the only difference is the presence of noncommutativity of the momentum sector $\bar{\theta}$, which splits the expected resonance point as $\varpi = \Omega \pm \Lambda_{0} = \Omega \pm \left( \Lambda_{\theta} - \Lambda_{\bar{\theta}} \right)$.  This also shows that if both the spatial and the momentum sector of the phase-space have a NC structure, their effects can not be separately identified, at least by looking at the resonance points. This answers the third question raised in the introduction.
Also notable is the existance of a smooth commutative limit $\left( \theta, \bar{\theta}\right) \to \left(0, 0 \right)$ i.e., $\left( \Lambda, \Lambda_{\bar{\theta}} \right) \to \left(0, 0 \right)$ when the resonance is at the expected frequency $\varpi = \Omega$.

Apart from affecting the resonance points, the effect of phase-space noncommutativity is also apparant in some terms in both (\ref{x11}, \ref{x22}) which are proportional to $\frac{f_{0}\Lambda}{4}$ and $\frac{4 f_{0}\Lambda_{\bar{\theta}}}{\varpi}$. We must try to estimate the size of such factors in context of HO system. Curiously, it turns out that they have a very diferent value in connection with the HO system than their corresponding estimate in the free particle context in the last section. To see this we have to consider the experimental setup where quantum mechanical HO's can be realized in GW detectors. We shall briefly elaborate on this after we consider the response of a HO system to circularly polarized GW in the next sub-section.

\subsection{Response to circularly polarized GW}
We shall now solve eqs.(\ref{e21aa}, \ref{e21bb}) for the circularly polarized GW.
We once again substitute the ansatz (\ref{e33},\ref{e43}) for the circularly polarized GW (\ref{e13}, \ref{e23}) in eqs.(\ref{e21aa}, \ref{e21bb})) to obtain the following set of coupled linear defferential equations
\begin{eqnarray}
\frac{dA}{dt} +i\varpi E + f_0\Omega C  +\frac{i4 \Lambda_{\bar{\theta}} }{ \varpi}f_0\Omega F + \Lambda_0 D &=& 0\>,
\nonumber \\
\frac{dB}{dt}   -\Omega C+i\varpi F-f_0\Omega D +\frac{i4 \Lambda_{\bar{\theta}} }{ \varpi}f_0\Omega E - \Lambda_0 C &=& 0\>,
\nonumber \\
\frac{dC}{dt}  + \Omega B +i\varpi G+ f_0\Omega A+\frac{i4 \Lambda_{\bar{\theta}} }{ \varpi}f_0\Omega H +\Lambda_0 B&=& 0\>,
\nonumber \\
\frac{dD}{dt} +i\varpi H - f_0\Omega B+\frac{i4 \Lambda_{\bar{\theta}} }{ \varpi}f_0\Omega G -\Lambda_0  A&=& 0\>
\nonumber \\
\frac{dE}{dt} +i\varpi A - f_0\Omega G -\frac{2i\Lambda_{\theta} }{\varpi }f_0\Omega B +\Lambda_0 H&=& 0\>,
\nonumber \\
\frac{dF}{dt} - \Omega G+i\varpi B + f_0\Omega H-\frac{2i\Lambda_{\theta} }{\varpi }f_0\Omega A -\Lambda_0 G &=& 0\>,
\nonumber \\
\frac{dG}{dt} + \Omega F +i\varpi C - f_0\Omega E -\frac{2i\Lambda_{\theta} }{\varpi }f_0\Omega D+\Lambda_0 F&=& 0\>,
\nonumber \\
\frac{dH}{dt} +i\varpi D + f_0\Omega F -\frac{2i\Lambda_{\theta} }{\varpi }f_0\Omega C-\Lambda_0 E &=& 0\>.
\label{iteration8}
\end{eqnarray}
\noindent Here $\Lambda_{\bar{\theta}}, \Lambda_{\theta}$ and $ \Lambda_{0}$ carry the same meaning as in the earlier sub-sections. Also note that the constant frequency $\Omega$ defined in (\ref{e13}, \ref{e23}) is the frequency of the circularly polarized GW. Solving the eq.(s)(\ref{iteration8}) to first order in the GW amplitude with boundary conditions (\ref{bc0}) which physically signifies that the GW hits the particle at t=0, we get 
\begin{eqnarray}
A(t) &=& 1 -\Lambda_0 V_2+\frac{\Lambda_0^2}{\Lambda_0^2-\varpi^2}-f_0\Omega V_3
+i\left[-\varpi V_1 - \frac{4\Lambda_{\bar{\theta}}f_0\Omega V_4}{\varpi}+\frac{4\Lambda_{\bar{\theta}}f_0\Omega}{\varpi}\frac{\left(\Omega-\Lambda_0\right)^2}{\varpi^2-\left(\Omega-\Lambda_0\right)^2}-\frac{\varpi^2}{\Lambda_0^2-\varpi^2}\right]\nonumber\\
B(t) &=& \left(\Omega+\Lambda_0\right)V_3-f_0\Omega V_2 + \frac{f_0\Omega\Lambda_0}{\Lambda_0^2-\varpi^2}
+i\left[-\varpi V_4+ \frac{4\Lambda_{\bar{\theta}}f_0\Omega V_1}{\varpi}+\frac{\varpi\left(\Omega-\Lambda_0\right)}{\varpi^2-\left(\Omega-\Lambda_0\right)^2}\right]\nonumber\\
C(t) &=& -\left(\Omega+\Lambda_0\right)V_4-f_0\Omega V_1+\frac{\Omega^2-\Lambda_0^2}{\varpi^2-\left(\Omega-\Lambda_0\right)^2}+\frac{f_0\Omega\varpi}{\Lambda_0^2-\varpi^2}
+i\left[-\varpi V_3-\frac{4\Lambda_{\bar{\theta}}f_0\Omega V_2}{\varpi}+\frac{4\Lambda_{\bar{\theta}}f_0\Omega\Lambda_0}{\varpi\left(\Lambda_0^2-\varpi^2\right)}\right]\nonumber\\
D(t) &=& f_0\Omega V_4 + \Lambda_0 V_1- \frac{f_0\Omega\left(\Omega-\Lambda_0\right)}{\varpi^2-\left(\Omega-\Lambda_0\right)^2}-\frac{\Lambda_0\varpi}{\Lambda_0^2-\varpi^2}
+i\left[-\varpi V_2 - \frac{4\Lambda_{\bar{\theta}} f_0\Omega V_3}{\varpi}+\frac{\varpi\Lambda_0}{\Lambda_0^2-\varpi^2}\right]\nonumber\\
E(t) &=& 1 -\Lambda_0 V_2+\frac{\Lambda_0^2}{\Lambda_0^2-\varpi^2}-f_0\Omega V_3
+i\left[-\varpi V_1 + \frac{2\Lambda_{\theta} f_0\Omega V_4}{\varpi}-\frac{\varpi^2}{\Lambda_0^2-\varpi^2}-\frac{2\Lambda_{\theta} f_0\Omega}{\varpi}\frac{\left(\Omega-\Lambda_0\right)}{\varpi^2-\left(\Omega-\Lambda_0\right)^2}\right]\nonumber\\
F(t) &=& \left(\Omega+\Lambda_0\right)V_3-f_0\Omega V_2 + \frac{f_0\Omega\Lambda_0}{\Lambda_0^2-\varpi^2}
+i\left[-\varpi V_4+\frac{\varpi\left(\Omega-\Lambda_0\right)}{\varpi^2-\left(\Omega-\Lambda_0\right)^2}+\frac{2\Lambda_{\theta} f_0 \Omega V_3}{\varpi}+\frac{2\Lambda f_0 \Omega}{\Lambda_0^2-\varpi^2}\right]\nonumber\\
G(t) &=& -\left(\Omega+\Lambda_0\right)V_4+f_0\Omega V_1+\frac{\Omega^2-\Lambda_0^2}{\varpi^2-\left(\Omega-\Lambda_0\right)^2}+\frac{f_0\Omega\varpi}{\Lambda_0^2-\varpi^2}
+i\left[-\varpi V_3 +\frac{2\Lambda_{\theta} f_0\Omega V_2}{\varpi}-\frac{2\Lambda_{\theta} f_0 \Omega}{\varpi}\frac{\Lambda_0}{\Lambda_0^2-\varpi^2}\right]\nonumber\\
H(t) &=& -f_0\Omega V_4 + \Lambda_0 V_1+ \frac{f_0\Omega\left(\Omega-\Lambda_0\right)}{\varpi^2+\left(\Omega-\Lambda_0\right)^2}-\frac{\Lambda_0\varpi}{\Lambda_0^2-\varpi^2}
+i\left[-\varpi V_2 +\frac{2\Lambda_{\theta} f_0 \Omega V_3}{\varpi}+\frac{\varpi\Lambda_0}{\Lambda_0^2-\varpi^2}\right]\nonumber\\
\label{100z}
\end{eqnarray}
where
$V_1$, $V_2$, $V_3$, $V_4$ are functions with dimensions of inverse frequency given by
\begin{eqnarray}
V_1&=& \frac{\left(\sin\varpi  t+\cos\varpi t\right)\left(\Lambda_0\sin\Lambda_0 t-\varpi\cos\Lambda_0 t\right)}{\Lambda_0^2-\varpi^2} \nonumber\\
V_2&=& \frac{\left(\sin\varpi  t+\cos\varpi t\right)\left(\Lambda_0\cos\Lambda_0 t+\varpi\sin\Lambda_0 t\right)}{\Lambda_0^2-\varpi^2} \nonumber\\
V_3&=&\frac{\varpi\cos\varpi t \sin\left(\Omega- \Lambda_0\right)t-\left(\Omega-\Lambda_0\right)\sin\varpi t \cos\left(\Omega-\Lambda_0\right)t}{\varpi^2-\left(\Omega-\Lambda_0\right)^2}\nonumber\\
V_4&=&\frac{\varpi\sin\varpi t \sin\left(\Omega- \Lambda_0\right)t+\left(\Omega-\Lambda_0\right)\cos\varpi t \cos\left(\Omega-\Lambda_0\right)t}{\varpi^2-\left(\Omega-\Lambda_0\right)^2}.
\label{100zhh}
\end{eqnarray}
In terms of the position and momentum expection values the solution for $\left(\langle X_{1}\left(t\right)\rangle, \langle X_{2}\left(t\right)\rangle\right)$ read
\begin{eqnarray}
\langle X_{1}\left(t\right)\rangle & = & \left[\left(1-\Lambda_0 V_2 + \frac{\Lambda_0^2}{\Lambda_0^2-\varpi^2}\right)+ \left(\Omega+\Lambda_0\right)\left(V_3\epsilon_3-V_4\epsilon_1\right)\right]X_{1}\left( 0 \right)\nonumber\\
&&+\left(\Omega+\Lambda_0\right)\left[\left(V_3\epsilon_1+V_4\epsilon_3\right)-\frac{\left(\Omega-\Lambda_0\right)}{\varpi^2-\left(\Omega-\Lambda_0\right)^2}\right]X_{2}\left( 0 \right)\nonumber\\
&&+\left[\left( V_1 - \frac{\varpi}{\Lambda_0^2-\varpi^2}\right)+\left(V_4\epsilon_3+V_3\epsilon_1\right)-\frac{\left(\Omega-\Lambda_0\right)\epsilon_3}{\varpi^2-\left(\Omega-\Lambda_0\right)^2}\right]\frac{P_{1}\left( 0 \right) }{m}+\left[\left(V_4\epsilon_1-V_3\epsilon_3\right)\right]\frac{P_{2}\left( 0 \right) }{m}\nonumber\\
&&+f_0\Omega\left[\left(V_3\epsilon_3-V_1\epsilon_1-V_3\right) -\frac{\varpi\left(\epsilon_3+\epsilon_1\right)}{\Lambda_0^2-\varpi^2}\right]X_{1}\left( 0 \right)\nonumber\\
&&+f_0\Omega\left[\left(V_2\epsilon_1+V_1\epsilon_3-2V_4\right)+\frac{2\left(\Omega-\Lambda_0\right)}{\varpi^2-\left(\Omega-\Lambda_0\right)^2}\right]X_{2}\left( 0 \right)\nonumber\\
&&-\frac{2f_0\Omega\Lambda_{\bar{\theta}}}{\varpi}\left[-2\left\{-V_4+\frac{\left(\Omega-\Lambda_0\right)}{\varpi^2-\left(\Omega-\Lambda_0\right)^2}\right\}-\left(V_1\epsilon_3-V_2\epsilon_1\right)+2\frac{\Lambda_0}{\Lambda_0^2-\varpi^2}\epsilon_1\right]\frac{P_{1}\left( 0 \right) }{m\varpi}\nonumber\\
&&-\frac{2f_0\Omega}{\varpi}\left[\Lambda_{\bar{\theta}}\left(V_1\epsilon_1+V_2\epsilon_3\right)+\left(\Lambda_{\bar{\theta}}+\Lambda_{\theta}\right)\left(V_3-\frac{\Lambda_0\epsilon_3}{\Lambda_0^2-\varpi^2}\right) \right]\frac{P_{2}\left( 0 \right) }{m\varpi}
\label{x1circular-h}
\end{eqnarray}
and 
\begin{eqnarray}
\langle X_{2}\left(t\right)\rangle & = & \left[\left(1-\Lambda_0 V_2 + \frac{\Lambda_0^2}{\Lambda_0^2-\varpi^2}\right)+ \left(\Omega+\Lambda_0\right)\left(V_3\epsilon_3-V_4\epsilon_1\right)\right]X_{2}\left( 0 \right)\nonumber\\
&&+\left(\Omega+\Lambda_0\right)\left[\left(V_3\epsilon_1+V_4\epsilon_3\right)+\frac{\left(\Omega-\Lambda_0\right)}{\varpi^2-\left(\Omega-\Lambda_0\right)^2}\right]X_{1}\left( 0 \right)\nonumber\\
&&+\left[\left( V_1 - \frac{\varpi}{\Lambda_0^2-\varpi^2}\right)+\left(V_4\epsilon_3+V_3\epsilon_1\right)-\frac{\left(\Omega-\Lambda_0\right)\epsilon_3}{\varpi^2-\left(\Omega-\Lambda_0\right)^2}\right]\frac{P_{2}\left( 0 \right) }{m}-\left[\left(V_4\epsilon_1-V_3\epsilon_3\right)\right]\frac{P_{1}\left( 0 \right) }{m}\nonumber\\
&&+f_0\Omega\left[\left(V_3\epsilon_3-V_1\epsilon_1-V_3\right) +\frac{\varpi\left(\epsilon_3+\epsilon_1\right)}{\Lambda_0^2-\varpi^2}\right]X_{2}\left( 0 \right)\nonumber\\
&&+f_0\Omega\left[\left(V_2\epsilon_1+V_1\epsilon_3-2V_4\right)-\frac{2\left(\Omega-\Lambda_0\right)}{\varpi^2-\left(\Omega-\Lambda_0\right)^2}\right]X_{1}\left( 0 \right)\nonumber\\
&&-\frac{2f_0\Omega\Lambda_{\bar{\theta}}}{\varpi}\left[2\left\{-V_4+\frac{\left(\Omega-\Lambda_0\right)}{\varpi^2-\left(\Omega-\Lambda_0\right)^2}\right\}-\left(V_1\epsilon_3-V_2\epsilon_1\right)+2\frac{\Lambda_0}{\Lambda_0^2-\varpi^2}\epsilon_1\right]\frac{P_{2}\left( 0 \right) }{m\varpi}\nonumber\\
&&+\frac{2f_0\Omega}{\varpi}\left[\Lambda_{\bar{\theta}}\left(V_1\epsilon_1+V_2\epsilon_3\right)-\left(\Lambda_{\bar{\theta}}+\Lambda_{\theta}\right)\left(V_3-\frac{\Lambda_0\epsilon_3}{\Lambda_0^2-\varpi^2}\right) \right]\frac{P_{1}\left( 0 \right) }{m\varpi}
\label{x2circular-h}
\end{eqnarray}
Once again we look at these solutions critically to point out their salient features. 

First, note that like in the previous cases, we have expressed the present solutions in a suggestive way where the first four terms represent the oscillatory nature of the HO system independent of any GW effect (these survive when there is no incoming GW, i.e. $f_{0} = 0$) whereas the last four terms manifestly show the effect of GW. 
A marked difference in the present case with the earlier ones is the existence of two distinct resonant points at 
\begin{eqnarray}
\varpi &=& \pm \Lambda_{0} \nonumber \\
\varpi &=& \Omega \pm \Lambda_{0}
\label{reso}
\end{eqnarray}
The first one correspond to a resonance between the natural frequency of the HO and phase-space characteristic frequency $\Lambda_{0}$ whereas the second one is the usual resonance between the GW and the HO system, although shifted by $\Lambda_{0}$. This result is again consistent with our earlier work \cite{ncgw4} where we considered response of HO to circularly polarized GW in a spatial NC background. Both the resonances were present in \cite{ncgw4}, however instead of the characteristic frequency $\Lambda_{0}$ of the NC phase-space only $\Lambda_{\theta}$ from the spatial NC sector appeared there. 

This, along with the result of the previous sub-section, answer all three questions posed in the introduction. It tells us that generalizing from the spatial noncommutativity to the phase-space noncommutativity  does not alter the basic features of the response of matter system to linearly or circularly polarized GW. That, as far as the resonance behaviour is concerned the momentum NC scale introduces a new frequency that combines with the corresponding characteristic frequency of spatial NC scale in such a way that looking at the resonance point their effects cannot be distinguished.

As in the earlier cases the present solution also exhibits the existence of a smooth commutative limit $\left( \Lambda, \Lambda_{\bar{\theta}} \right) \to \left(0, 0 \right)$. Note that here the first resonance point ceases to exist and the second one reduces to the usual case $\varpi = \Omega$ when we go to the commutative limit. 

Similar to the previous section, here also some terms appear with  coefficients $\frac{2f_0 \Omega  \Lambda_{\bar{\theta}}}{\varpi}$ and $\frac{2f_0 \Omega \Lambda_{\theta}}{\varpi}$, which may or may not be of significant size in experimental realization of the quantum mechanical HO in context of GW detectors. We estimate the size of such terms using the currently available upper-bounds of the NC parameters $\theta$ \cite{carol} and $\bar{\theta}$ \cite{bert0, samanta} in the next sub-section.
\subsection{Realization of quantum mechanical HO in Resonant Bar Detectors}
At present there are mainly two types of GW detectors that are operational, one is the ground-based interferometric detector which operates in a wide frequency spectrum and the other is the resonant bar detector operating in a relatively small frequency window. Both these detectors operate in the long wavelength and low velocity regime where our calculation in the present paper applies. Specifically the HO system that has been considered in the last two sub-sections is readily realized in context of the resonant bar detectors pioneered by J.Weber \cite{Weber, Resonant_Bar_Review}. These are typical cylindrical Aluminium bars of length $L \sim 3 \,{\rm m}$ and width $r = 30\, {\rm cm}$, weighing nearly $M = 2 \times 10^{3}\, {\rm Kg}$. For such bar-detectors it can be shown \cite{Magg} that the fundamental mode of elastic oscillation driven by the passing GW is identical to a forced harmonic oscillator with effective mass $m_{0} = \frac{M}{2} \sim 10^{3}\, {\rm Kg}$ and frequency $\varpi_{0} \sim \frac{\pi v_{s}}{L} = 5.6 \,{\rm kHz}$, where $v_{s} \sim 5.4\, {\rm km \,sec^{-1}}$ is the sound speed within the Aluminium bar at low temperature \footnote{Let us also note that at (near) the resonance the typical reduced wavelength of the GW $\lambdabar = \frac{c}{\varpi_{0}}$ will be such that $\frac{L}{\lambdabar} = \frac{\pi v_{s}}{c} = 6 \times 10^{-5} << 1$ ensuring that the long wave-length, low-velocity limit is satisfied. }. Thus the bar detector, despite being a $2$-ton macroscopic object, responds to the GW by generating collective elastic oscillation modes that are so tiny in size that a quantum mechanical treatement is necessary and upon quantization they are called the phonon modes. Our quantum mechanical analysis of the HO system in this section applies to these phonon modes. 

Using the existing upper-bounds on the spatial NC parameter \cite{carol} $\theta \sim 10^{-40} {\rm m}^{2}$ the corresponding NC frequency $\Lambda_{\theta}$ 
for such a HO mode with KHz-range frequency and $m_{0} \sim 10^{3}\, {\rm Kg}$ effective mass will be 
\begin{equation}
\Lambda_{\theta} \sim  \left(\frac{m}{m_{0}} \right) \left(\frac{\varpi}{1 {\rm KHz}}\right)^{2} \left(\frac{\theta}{ 10^{-40} {\rm m}^{2}}\right)  \times 0.5 \,{\rm KHz}
\label{sp_nc_f}
\end{equation}
i.e., typically in the KHz range. So it will significantly alter the resonance point. But for the same HO mode the characteristic frequency $\Lambda_{\bar{\theta}}$ set by the momentum NC scale \cite{bert0, samanta} will be  
\begin{eqnarray}
\Lambda_{\bar{\theta}} \sim \left(\frac{\bar{\theta}}{2.32 \times 10^{-61} {\rm Kg}^{2} {\rm m}^{2} {\rm sec}^{-2}}\right) \left(\frac{m_{0}}{m} \right) 0.5 \times 10^{-30} \,{\rm Hz}
\label{m_nc_f}
\end{eqnarray}
which is way beyond the operating range of the bar-detectors, and thus can not affect the resonance point. Thus we conclude that though in general the results in the HO-GW interaction case will depend on the composit NC phase-space frequency $\Lambda_{0}$, but for the HO modes in a typical bar detectors $\Lambda_{0} \approx \Lambda_{\theta}$.
The dimention-less parameter $\Lambda$ in the present case takes value of the order of unity
\begin{equation}
\Lambda = \frac{m\varpi \theta}{\hbar} \approx 1 \times \left(\frac{m}{m_{0}}\right) \left(\frac{\varpi}{1 {\rm KHz}} \right) \left( \frac{\theta}{10^{-40} {\rm m^{2}}}\right)
\label{dim_less_p}
\end{equation}
So in case of linearly polarized GW, terms in the solutions (\ref{x11}, \ref{x22}) containing the factor $\frac{f_{0}\Lambda}{4}$ are ${\cal{O}}\left[ f_{0}\right]$. They will grow rapidly near the resonance points and affect the system significantly. But terms proportional to $\frac{4 f_{0}\Lambda_{\bar{\theta}}}{\varpi}$ will be ${\cal{O}}\left[ f_{0} \times 10^{-33}\right]$, so even near resonance they will be much smaller than the other terms. 

Since the bar detector operates in the KHz range, only GW's with similar frequency is of concern here. Thus $\left(\frac{\Omega}{\varpi}\right) \sim {\cal{O}}\left[ 1\right]$ and terms with coefficients such as $\frac{2f_0 \Omega \Lambda_{\theta}}{\varpi}$ and $\frac{2f_0 \Omega  \Lambda_{\bar{\theta}}}{\varpi}$, appearing in the solution (\ref{x1circular-h}, \ref{x2circular-h}) for circularly polarized GW are ${\cal{O}}\left[ f_{0}  \Lambda_{\theta}\right]$ and ${\cal{O}}\left[ f_{0}  \Lambda_{\bar{\theta}}\right]$ respectively. It is evident from equations (\ref{sp_nc_f}, \ref{m_nc_f}) and the solution (\ref{x1circular-h}, \ref{x2circular-h}) that while terms with the formar coefficient are ${\cal{O}}\left[ f_{0} \right]$ and will grow significant near resonance, those with the latter coefficients are ${\cal{O}}\left[ f_{0} \times 10^{-33} \right]$ and will remain sub-dominant in comparison. So in both the cases of linearly and circularly polarized GW, the momentum NC does not affect the result much as far as the bar detectors are concerned. However, in general the response of a HO system to GW depends on momentum NC parameter and such dependence may become significant for some other realizations of quantum HO systems as GW detectors. 
%
\section{Concluding remark}
%
In the present paper we have considered how a free particle and harmonic oscillator (HO) in a quantum domain will respond to linearly and circularly polarized gravitational waves (GW) if the given phase-space has a noncommutative (NC) structure. The results show resonance behaviour in the responses of both free particle and HO systems to GW with both kind of polarizations. 

While HO system with a natural frequency is expected to display resonance with GW of suitable frequency, it is rather curious that a free particle may also show resonance. In our analysis it turns out that momentum noncommutativity induces a oscillatory behaviour in the free particle with a characteristic frequency $\Lambda_{\bar{\theta}}$ and that in turn can resonate with the GW of both polarizations, but spatial noncommutativity does not affect the response in any detectable way. Thus it may be possible to test the existence of noncommutativity in the momentum sector of the phase-space algebra by looking at how free particle respondes to GW. Any resonance in the response is an evidence in favour of such noncommutativity. Also it should be noted that even in absence of any GW the free particle's response still contain oscillatory terms due to momentum noncommutativity. This will show up as a oscillatory noise with characteristic NC frequency $\Lambda_{\bar{\theta}}$ in any GW detector made based on free-particle dynamics. 

In the response of HO system (with natural frequency $\varpi$) to GW (with frequency $\Omega$), we found that both the spatial and the momentum sector of the noncommutative phase-space algebra introduce two different characteristic frequencies ($\Lambda_{\theta}$ and $\Lambda_{\bar{\theta}}$ respectively) into the system, but these two frequencies combine together to form a composit frequency $\Lambda_{0} = \left( \Lambda_{\theta} - \Lambda_{\bar{\theta}} \right)$ which splits the expected resonance point from $\varpi = \Omega$ to  $\varpi = \Omega \pm \Lambda_{0}$.
So two equally spaced resonance points instead of a single one at the centre is an evidance for a noncommutative phase-space structure. There is also another resonance point at $\varpi = \pm \Lambda_{0}$ between the frequency $\Omega$ of the rotating triad and the NC phase-space frequency $\Lambda_{0}$ which can be a false alarm for GW detection that one should be cautious about. 
 
From our calculations of the quantum mechanical free particle case we saw that if free neutrons can be subjected to the GW, the corresponding $\Lambda_{\bar{\theta}}$ will be in the Hz-range, whereas the dimensionless parameter $\Lambda$ will be a very small number. Also the spatial noncommutativity will not affect the neutron's response in any measurable way.

On the other hand, we have argued that the presently operating resonant bar detectors of GW are experimental realization of the quantum HO systems we have considered in the present paper. In this case the effective mass of the fundamental phonon mode (phonons are quantized vibrational modes of the bar detector that behave as HO) is $\sim 10^{3} {\rm Kg}$ \cite{Magg} and the dimensionless parameter $\Lambda$ is approximately unity, spatial NC frequency $\Lambda_{\theta}$ is in KHz range and momentum NC frequency $\Lambda_{\bar{\theta}}$ is far below the Hz-range. These numbers are in complete contrast with the earlier free particle case. 
We thus conclude that the mass (effective or otherwise) of a test body that is subjected to the GW, plays a crucial role to determine the relative size of the NC parameter dependent terms in various solutions and hence different NC dependent response terms may become important in context of different realization of the quantum free particle or harmonic oscillator in various GW detection scenario.
   
\section*{Acknowledgemnet} 
AS acknowledges the finantial support of DST SERB under Grant No. SR/FTP/PS-208/2012. SG acknowledges the finantial support of DST SERB under Grant No. YSS/2014/000180.

\end{document}